\documentclass[12pt,a4paper]{article}
\usepackage[utf8x]{inputenc}
\usepackage{jheppub}
\usepackage{youngtab}
\usepackage{graphicx}
\newcommand{\one}{{\rm 1\kern -.9mm l}}

\title{\center{Gauge-Stringy Instantons in $\mathcal N=2$ $U(N)$ Gauge Theories}}

\author{Hossein Ghorbani }

\affiliation{School of Particles and Accelerators,\\ Institute for Research in Fundamental
Sciences (IPM),\\
P.O. Box 19395-5531, Tehran, Iran}

\emailAdd{pghorbani@ipm.ir}

\abstract{Using D3/D(-1) brane set-up in type IIB string theory we introduce gauge-stringy instantons in $\mathcal N=2$ U$(N)$
supersymmetry theories with one matter multiplet in symmetric representation.
In addition to the gauge and stringy moduli there exist extra zero modes
that we refer to as ``gauge-stringy'' moduli. We show that the measure of the moduli space in this model becomes dimensionless 
for arbitary $N$ when the gauge instanton charge $k_g$ is equal to the stringy instanton charge $k_s$. 
This property of gauge-stringy instantons leads to having equal contributions from all instanton charges $k_s=k_g\equiv k$ in the effective action. 
We derive the gauge-stringy instanton partition function and calculate the corrections to the prepotential due to $k=1,2$ 
gauge-stringy instanton charges. As a by-product the partition function for gauge $k$-instanton is obtained which coincides
with the result from the standard ADHM construction.
}


\begin{document}
\maketitle 

\section{Introduction}

Exploiting the localization technique pioneered by N. Nekrasov \cite{Nekrasov:2002qd}, multi-instanton calculus 
in supersymmetry field theories in recent years has received
remarkable achievements (see for instance \cite{Dorey:2002ik,Bianchi:2007ft}).
On the other hand, string theory beside giving a realization of the gauge instantons \cite{Witten:1995gx,Douglas:1995bn,Billo:2002hm} 
has also revealed new features of non-perturbative effects through the introduction of the so-called stringy instantons 
\cite{Argurio:2007vqa,Florea:2006si}.
In string theory the instantonic configurations can be engineered by D$p$/D$(p-4)$ brane set-ups. 
For the cases that $p$ is greater than three one needs to compactify along some directions of the brane world-volume
in order to have supersymmetry gauge theories in four dimensions. 
The simplest example that does not need any compactifications in these group of set-ups is D3/D(-1) system that we will consider thoughout this article. 
Having picked the D3/D(-1) brane set-up, the $\mathcal N=2$ gauge theories which is of our interest in this work is provided when 
among other mechanisms we reduce the number of supersymmetries (from $\mathcal N=4$ living on D3 branes) by
including an orbifold in the background. In the presence of orbifold the fractional branes are defined \cite{Franco:2005zu,Diaconescu:1997br}, 
and though their existence there will be geometrically different arrangements of D3 and D(-1) 
branes which lead to non-trivial gauge and stringy non-perturbative effects in the effective action of the theory. 

There are some works in the literature that study the stringy instanton effects in the low-energy effective actions
(see e.g. \cite{Billo':2010bd,Ghorbani:2010ks,Petersson:2007sc}) 
and some of their applications from the phenomenological point of view \cite{Ibanez:2008my,Kiritsis:2009sf,Blumenhagen:2007zk,Cvetic:2007ku}. 
It turns out that stringy instanton corrections in the effective action 
from instantons with charge $k$
are suppressed by powers of string length scale $\alpha'$ when the instanton charge $k$ increases \cite{Ghorbani:2010ks}. There are some exceptions that 
effective action corrections do not depend on $\alpha'$ even if the corrections stem from a completely stringy nature. One example 
is given in \cite{Ghorbani:2010ks} where the stringy instanton effects in the effective action of $\mathcal N=2$ SU($2$) with a 
matter hypermultiplet in symmetric representation has been calculated. In that model although D-instantons posses 
a pure stringy characteristic, for one special gauge group i.e. SU$(2)$ the corrections lose the dependency on the string parameter $\alpha'$. 
It is known that this certain theory has a vanishing one-loop coefficient of the $\beta$-function (and the only $\beta$-function 
coefficient due to the non-renormalization theorem for $\mathcal N=2$ supersymmetry \cite{Seiberg:1988ur}) for the group SU$(2)$. The reason that in the special 
model used in \cite{Ghorbani:2010ks}, stringy instantons contribute equally in all orders in the effective action, is that the 
measure of the moduli space happens to be dimensionless for the gauge group SU$(2)$.
One is then allowed to sum up the contributions from all instanton charges leading to an exact determination of the effective action from stringy corrections. 
In \cite{Ghorbani:2010ks} the stringy corrections in the prepotential function was calculated up to $k=5$ instanton charge. From the form of the corrections 
it was easy to conjecture that corrections due to different instanton charges are the expansion terms of a closed form given by:
\begin{equation*}
 F^{(n.p.)}(\Phi)=-\text{Tr}\Phi^2 \log \left(1+\frac{\mathcal N_1}{2} q \right) 
\end{equation*}
where $\mathcal N_1$ is the normalization constant for 1-instanton partition function and $q\equiv \exp(2\pi \tau)$ with $\tau$ being 
the complex coupling constant, is a dimensionless quantity. 
Using a relation between the elementary symmetric polynomials and power sums (see for instance \cite{Macdonald:1995})
one is able to generalize the results in \cite{Ghorbani:2010ks} to the gauge group SU$(N)$ \cite{Ghorbani:2011xh}. 
However for a generic SU$(N)$ gauge theory the stringy corrections in the prepotential will depend on the parameter $\alpha'$. 

A question that may arise is that under what conditions we may obtain a closed form for the instanton corrections in the effective action 
in all orders of instanton charges for a general gauge group SU$(N)$ or even other supersymmetry gauge theories. 
This requires having a dimensionless moduli measure for the moduli space of the given theory. 
In the current work we try to answer this question by introducing a model of the same D3/D(-1) system that we explored in \cite{Ghorbani:2010ks,Ghorbani:2011xh}.
Our answer therefore will be suitable only for $\mathcal N=2$ U$(N)$ gauge theories with one matter in symmetric representation. However similar 
approach may be taken for other supersymmetry gauge theories. 

The model we use here is simply the combination of gauge and stringy instanton brane set-ups in type IIB. One may expect that zero modes
in the theory consists of only the combination of the gauge instanton and stringy instanton zero-modes, but in fact it turns out that in addition to these moduli
there exist extra zero-modes in the spectrum that we refer to as \emph{gauge-stringy neutral moduli}. 
In the language of the states in the D3/D(-1) brane set-up, the new zero-modes correspond to 
bosonic and fermionic neutral string states stretching between D(-1) branes sitting in two different representations of the orbifold group.
The main issue we concern in this paper is regarding the role of the gauge-stringy moduli in the measure of the moduli space and their contribution in the 
instanton partition function.  

The content of this paper is as the following: In the next section we describe the model in the D3/D(-1) brane set-up. In section \ref{sec-spectrum} the 
spectrum of the moduli space and their Chan-Paton structure are discussed. We also define the auxiliary moduli that are necessary in writing the moduli action as 
a BRST $Q$-exact action. In section \ref{sec-action} we present the full action for the current model. The action can be decomposed into different parts of
gauge, stringy and gauge-stringy actions. In section \ref{sec-dimless} we show that under some conditions the measure of the moduli space becomes dimensionless. 
Then in section \ref{sec-patition} we turn into calculating the instanton partition functions for different types of instantons, in particular the gauge-stringy
partition function. Finally in section \ref{sec-pre} we calculate the prepotential corrections from gauge-stringy instantons with charges $k=1,2$. In 
\ref{sec-con} we conclude with a summary of the results and a discussion on the prospective of the future work.

\section{Gauge-Stringy Set-up}\label{sec-review}

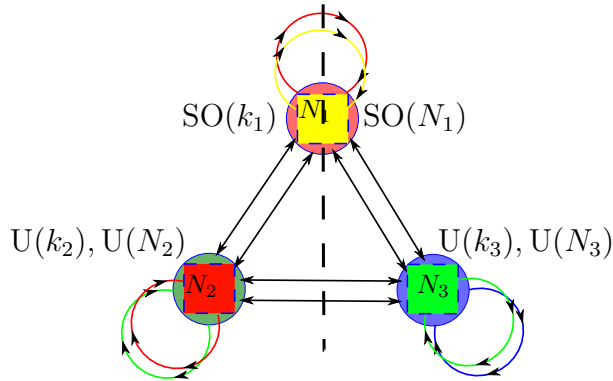
\begin{figure}[hbt]\label{QGSD}
  \begin{center}
\vspace{10pt}
\hspace{-30pt}
\begin{picture}(0,0)%
\includegraphics[scale=0.26]{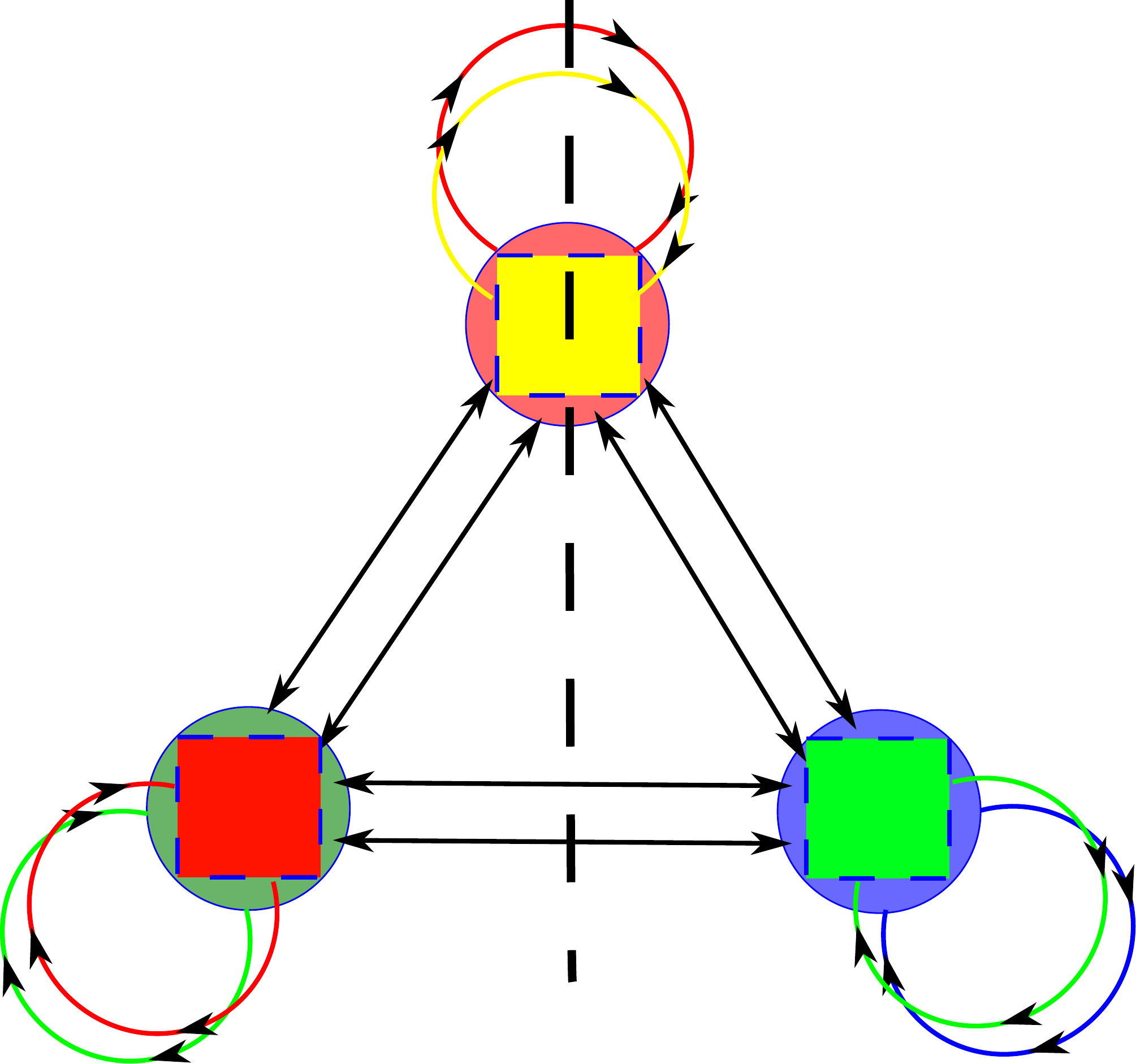}%
\end{picture}%
\setlength{\unitlength}{1450sp}%
\begingroup\makeatletter\ifx\SetFigFontNFSS\undefined%
\gdef\SetFigFont#1#2#3#4#5{%
  \reset@font\fontsize{#1}{#2pt}%
  \fontfamily{#3}\fontseries{#4}\fontshape{#5}%
  \selectfont}%
\fi\endgroup%
\begin{picture}(6030,5737)(541,-4460)
\put(1651,-2281){\makebox(0,0)[rb]{\smash{{\SetFigFont{12}{9.6}
 {\familydefault}{\mddefault}{\updefault}$\mathrm{U}(k_2),\mathrm{U}(N_2)$}}}}
\put(2181,-2990){\makebox(0,0)[rb]{\smash{{\SetFigFont{10}{9.6}
 {\familydefault}{\mddefault}{\updefault}$N_2$}}}}

 \put(6400,-151){\makebox(0,0)[rb]{\smash{{\SetFigFont{12}{9.6}
 {\familydefault}{\mddefault}{\updefault}$\mathrm{SO}(N_1)$}}}}
 \put(3191,-151){\makebox(0,0)[rb]{\smash{{\SetFigFont{12}{9.6}
 {\familydefault}{\mddefault}{\updefault}$\mathrm{SO}(k_1)$}}}}
 
 \put(8891,-2281){\makebox(0,0)[rb]{\smash{{\SetFigFont{12}{9.6}
 {\familydefault}{\mddefault}{\updefault}$\mathrm{U}(k_3),\mathrm{U}(N_3)$}}}}

 \put(6111,-2990){\makebox(0,0)[rb]{\smash{{\SetFigFont{10}{9.6}
 {\familydefault}{\mddefault}{\updefault}$N_3$}}}}

 \put(4111,-10){\makebox(0,0)[rb]{\smash{{\SetFigFont{10}{9.6}
 {\familydefault}{\mddefault}{\updefault}$N_1$}}}}
\end{picture}%
  \end{center}
 \label{Fig:1}
\caption{Quiver diagram for the most general gauge-stringy instanton configuration.}
 \end{figure}

The D3/D(--1) set-up consists of $N$ D3-branes extended along first four coordinates of 10-dimensional space-time together 
with $k$ D(--1) branes localized in all space-time directions. Table 1. shows the boundary conditions on string endpoints
in the presence of D3/D(--1) branes. The stack of D3-branes on its own leads to $\mathcal N=4$ U(N) 
gauge theory in the low-energy regime. 
Our goal in this work is to study different types of instantons in $\mathcal N=2$ supersymmetries. We therefore need to reduce 
somehow the number of supersymmetries. One way to break supersymmetry is to invoke
extra symmetries in certain directions of space-time. In other words,
we may reduce the number of supersymmetries by means of orbifold background in the space-time.  
In order to study instantons in $\mathcal{N}=2$, following the model described in \cite{Ghorbani:2011xh,Ghorbani:2010ks} 
we act the orbifold group $\mathbb Z_3=\{1,\xi,\xi^{-1}\}$ with $\xi=e^{2\pi i/3}$ on the first two complex internal coordinates,
so that the geometry of the internal space becomes 
$\mathbb C^2 / \mathbb Z_3 \times \mathbb C$; the third complex coordinate in the internal space therefore 
remains unchanged by orbifolding. In addition to the orbiforld background in the space-time we project out further 
string states by including orientifold O3-plane in the background. If the orientifold plan lies along the same directions as D3-branes,
no more supersymmetry breaking takes place, instead some degrees of freedom discarded out from the spectrum. 
It is elaborated in \cite{Argurio:2007vqa} that it is exactly the absence of these states that leads to the so-called stringy instantons. 
The Chan-Paton matrices of the string excitations on D-branes/D-instantons can now transform in three ways corresponding to three 
representations of the orbifold group $\mathbb Z_3$. $N$ D3-branes ($k$ D-instantons) split into three sets of 
$N_1,N_2$ and $N_3$ fractional D3-branes
($k_1,k_2$ and $k_3$ fractional D-instantons) each being in one representation of the orbifold group. The gauge/instanton groups living on
three different sets of branes are schematically illustrated by \emph{gauge-quiver} diagram in \figurename { 1}. The quiver for 
instantons on $\mathbb{C}^2/\mathbb{Z}_3$ was proposed in \cite{Intriligator:1997kq} (The idea of quiver for instantons in ALE spaces
was introduced earlier by Douglas and Moore \cite{Douglas:1996sw}).

In the gauge-quiver language the gauge 
instanton is the configuration of sitting D3-branes and D-instantons on the same node of the quiver diagram, say node 2. The stringy instanton on the other hand 
is the configuration that
D3-branes and D-instantons lie on two different nodes of the quiver diagram, e.g. D-instantons on node 1 and D3-branes on node 2. The most general case 
is obviously when all the nodes are occupied by D3-branes/D-instantons (see Figure 1). Let us remind that due to the existence of the O3-plane along 
the four first coordinates there will be an identification between gauge groups U$(N_2)$ and U$(N_3)$ on nodes 2 and 3. The same happens for the 
gauge instanton groups U$(k_2)$ and U$(k_3)$. The gauge group U$(N_1)$ on node 1 after 
orientifolding reduces to O$(N_1)$. As we are interested in studying the instantons in U$(N)$ gauge theories we assume only $N$ D3-branes being on
node 2 (and the same numbers of them on node 3 because of orientifolding) and no such gauge branes on node 1. However to see the effect 
of both gauge and stringy instantons in the same time we consider $k_s$ D-instantons on node 1 and $k_g$ D-instantons on node 2 
(and equivalently on node 3). \\
The string states living only on D3-branes on node 2 constitute the pure $\mathcal N=2$ U$(N)$ supersymmetry. The strings which stretch
between D3-branes on nodes 2 and 3 make a matter hypermultiplet in the symmetric representation of the gauge group.
The details about the supersymmetry content in this model can be found in \cite{Ghorbani:2010ks}. In the next section we study in detail the spectrum of the gauge-stringy
configuration.

\begin{table}[ht]
\begin{center}
\begin{tabular}{c|cccc|cccccc}
\phantom{\vdots}
& 0 & 1& 2&3 &4& 5& 6&7 &8 &9 
\\
\hline
\phantom{\vdots}D3&$-$ &$-$&$-  $&$-$&$ \times$&$\times$&$\times$&$\times$&$\times$&$\times$\\
\hline
\phantom{\vdots}D(--1)&$\times$&$\times$&$\times$&$\times$
&$\times$&$\times$&$\times$&$\times$&$\times$&$\times$\\
\end{tabular}
\end{center}
\label{tab-d3d-1}
\caption{The symbols $-$ and $\times$ respectively denote 
the Neumann and Dirichlet boundary conditions that the string endpoints on D3 or D(-1) branes satisfy. }
\end{table}

\section{Gauge-Stringy Moduli Spectrum}\label{sec-spectrum}

The moduli spectrum in the D3/D(--1) system stem from all string states living solely on D(--1)-branes or the string excitations 
stretching between D3-branes and D(--1)-branes.
In our model with the arrangement ($k_s$,$k_g$) for D-instantons and ($0$,$N$) for gauge branes, the moduli are divided into different classes. 
They are stringy (gauge) neutral moduli corresponding to the states arisen from D-instantons at node 1 (2), and the \emph{gauge-stringy neutral moduli}  
which are characterized by the states between D-instantons on nodes 1 and 2. The stringy (gauge) charged moduli are the states between D-instantons
on node 1 (2) and the gauge branes on node 2. 

The Chan-Paton factors are $3$ by $3$ block matrices. The entry $(ij)$ of a Chan-Paton matrix represents the string state with an endpoint attached to 
gauge brane or D-instanton on node $i$ and another endpoint attached to gauge brane or D-instanton on node $j$.

\subsection{Neutral Bosonic Moduli}

The NS sector of the string state on D-instanton is a boson in ten dimensions. We conventionally split the boson into one part along the D3-brane world-volume
and another part transverse to it, i.e. 
\begin{equation}
\phi_{M}\rightarrow a_{\mu}\oplus\chi_{p}
\end{equation}
where $M=0,...,9$; $\mu=0,...,3$ and $p=4,...,9$. Instead of the six real scalars in the internal space we work with three complex scalars because their
transformations under orbifold group is simpler. 
We define the complex scalars as $\chi_1\equiv\chi_4+i\chi_5$, 
$\bar\chi_2\equiv\chi_6-i\chi_7$, $\chi\equiv\chi_3\equiv\chi_8+i\chi_{9}$ and so on.
The bosonic moduli $a_\mu$ and the complex scalars $\chi_i$ with $i=1,2,3$, get affected differently from the orbifold and orientifold 
actions. Taking into account the consistency condition that guarantees the commutativity of the orbifold and orientifold 
transformations \cite{Gimon:1996rq},
neutral bosonic moduli must satisfy the following relations: 
\begin{subequations}\label{achi}
\begin{align}
& a_\mu =\gamma(g)\; a_\mu\; \gamma(g)^{-1} \;\;\;\;\;\;\;\;\; && a_\mu =\gamma_+(\Omega)\; a_\mu^T \; \gamma_+(\Omega)^{-1}\\
& \chi_i=\xi^i\; \gamma(g)\; \chi_i \; \gamma(g)^{-1}\;\;\;\; && \chi_i=-\gamma_+(\Omega) \; \chi_i^T \; \gamma_+(\Omega)^{-1} 
\end{align}
 \end{subequations}
where 
\begin{equation}\label{gamma}
 \gamma(g) = 
\begin{pmatrix}
\one_{k_s}&0&0\cr  
0&\xi\,\one_{k_g}&0\cr
0&0&\xi^{-1}\,\one_{k_g}
\end{pmatrix}
\;\;\;\;\;\;
\gamma_+(\Omega) = 
\begin{pmatrix}
\one_{k_s}&0&0\cr  
0&0&\one_{k_g}\cr
0&\one_{k_g}&0
\end{pmatrix}
\end{equation}
being $3\times 3$ block matrices, are the representations of the orbifold and orientifold groups acting on moduli Chan-Paton matrices. $k_s$ and $k_g$ are respectively the number 
of D-instantons on node 1 and 2, $g\in \mathbb Z_3$ and 
$\Omega=\omega\;(-1)^{F_L}\;\mathcal I_{456789}$ is the orientifold operator acting on string states. 
Note that the matrix representation of the orientifold group in (\ref{gamma}) acts only on the endpoint of the string in the Chan-Paton matrix that is attached to
D-instantons. The matrix representation of the orientifold group which acts on the string endpoint sitting on gauge branes is different from that of (\ref{gamma})
and is given in subsection (\ref{chargedm}) where we discuss the charged moduli. 
The bosonic Chan-Paton matrices satisfying the conditions (\ref{achi}) take the following forms:
\begin{equation}
 a^\mu = \left(\begin{array}{ccc}
               a^{\mu}_{(s)} & 0 & 0\\
               0 & a^{\mu}_{(g)} & 0\\
               0 & 0 & {a^{\mu}_{(g)}}^T
              \end{array}\right) ~\ \ \ \ \ \ \ \ \ 
\chi_3\equiv\chi = \left(\begin{array}{ccc}
               \chi_{(s)} & 0 & 0\\
               0 & \chi_{(g)} & 0\\
               0 & 0 & {-\chi_{(g)}}^T
              \end{array}\right) ~
\end{equation}
with $a^\mu_{(s)}={a^\mu_{(s)}}^T$, $\chi_{(s)}={-\chi_{(s)}}^T$, and
\begin{equation}\label{chi1chi2}
\chi^{1}=\left(\begin{array}{ccc}
0 & \chi^{1}_{(gs)} & 0\\
0 & 0 & \chi^{1}_{(g)}\\
-{\chi^{1}_{(gs)}}^T & 0 & 0
\end{array}\right)~\ \ \ \ \ \ \ \ \ \ \ \ 
 \chi^{2}=\left(\begin{array}{ccc}
0 & 0 & \chi^{2}_{(gs)}\\
{-\chi^{2}_{(gs)}}^T & 0 & 0\\
0 & \chi^{2}_{(g)} & 0
\end{array}\right)
 \end{equation}
where $\chi^{1}_{(g)}={-\chi^{1}_{(g)}}^T$ and $\chi^{2}_{(g)}={-\chi^{2}_{(g)}}^T$.
The subscripts ``g'', ``s'', and ``gs'' denote respectively the gauge, stringy and gauge-stringy characteristics of the moduli in the entries of the Chan-Paton
matrices. The instanton symmetry group is the product of the stringy and gauge instanton groups, i.e. SO$(k_s)\times$U$(k_g)$. The moduli $a^\mu_{(s)}, 
\chi_{(s)}$
and $a^\mu_{(g)}$, $\chi_{(g)}$ are respectively in the adjoint representations of the groups SO$(k_s)$ and U$(k_g)$. The moduli $\chi^1_{(g)}$ 
and $\chi^2_{(g)}$ are in the 
antisymmetric representation of the group U$(k_g)$ while $\chi^1_{(gs)}$ and $\chi^2_{(gs)}$ are in the bi-fundamental representation of the groups
SO$(k_s)$ and U$(k_g)$.
\subsection{Neutral Fermionic Moduli}

Similar to the neutral bosonic moduli, in the presence of D3-branes, the spinors in 10-dimensions are also split into spinors in directions 
along and transverse to the world-volume of D3-branes:
\begin{equation}
\Lambda_{\mathcal{\dot A}} \to\lambda_{\dot\alpha A}\oplus M^{\alpha A}
\end{equation}
where $\Lambda_{\mathcal{\dot A}}$ with $\mathcal{\dot A}=1,...,16$ is the anti-chiral spinor in 10d, $\alpha, \dot\alpha = 1,2$ are the
left and right handed spinor indeces in the Lorentz space (D3-brane world volume),
and $A=1,..,4$ is the spinor index in the 6d internal space. The 10d chiral spinor is discarded due to the GSO projection in type II. 
The modulus $\lambda_{\dot\alpha A}$ ($M^{\alpha A}$) is anti-chiral in 10d because it is anti-chiral (chiral) in the Lorenz space 
and chiral (anti-chiral) in the 
internal space. Note that the orbifold group acts only on the first two complex coordinates in the internal space. The symmetry group associated
to the internal space is SO$(6)$ which is homomorphic to SO$(4)\times$SO$(2)\sim$SU$(2)_L\times$SU$(2)_R\times$SO$(2)$. The index $A$ 
therefore can be split into left and right handed SU$(2)$ group indices, $a, \dot a=1,2$, plus the SO$(2)$ helicities.

Let us remind that these moduli are again $3\times 3$ block matrices which transform under orbifold and orientifold groups.
The orbifold group acts only on the spinor index in the internal space and leaves the spinor in the Lorentz space invariant. 
The orientifold background put extra constraints on 
the entries of the Chan-Paton matrices. The Chan-Paton matrices of neutral fermionic moduli then should satisfy the following conditions:
\begin{subequations}\label{cond2}
 \begin{align}
  &M^{\alpha A}=R(g)^A_{~B} \gamma(g) M^{\alpha B} \gamma(g)^{-1}~~&&M^{\alpha A}=R(\Omega)^A_{~B} \gamma_+(\Omega) (M^{\alpha B})^T \gamma_+(\Omega)^{-1}\\
  &\lambda_{\dot\alpha A}=\gamma(g) \lambda_{\dot\alpha B} \gamma(g)^{-1} R(g)^B_{~A}~~~~
  &&\lambda_{\dot\alpha A}=\gamma_+(\Omega) (\lambda_{\dot\alpha B})^T \gamma_+(\Omega)^{-1} R(\Omega)^B_{~A}
 \end{align}
\end{subequations}
 where $R(g)=e^{\frac{2\pi i}{3} J^{45}} e^{\frac{-2\pi i}{3} J^{67}}$ with $J^{45}$ and $J^{67}$ being the rotation generators along the 
 first two complex coordinates,
 is the orbifold action in the spinor representation. $\gamma(g)$ is the orbifold matrix representation
 acting on the Chan-Paton matrices. The orientifold acts only on the internal space spinors through $R(\Omega)=-\Gamma^{456789}$ which is a reflection 
 in all internal coordinates and reveals the chirality of the spinors in the internal space. Using $\gamma(g)$ and $\gamma_+(\Omega)$ given in
 (\ref{gamma}) the fermionic neutral moduli which satisfy the conditions (\ref{cond2}) have the following Chan-Paton forms:
\begin{subequations}
\begin{align}
& M^{\alpha \dot{a}}=
\left(\begin{array}{ccc}
M_{(s)}^{\alpha \dot{a}} & 0 & 0\\
0 & M_{(g)}^{\alpha \dot{a}} & 0\\
0 & 0 & {M_{(g)}^{\alpha \dot{a}}}^T
\end{array}\right)
&& \lambda_{\dot{\alpha} \dot{a}}=
\left(\begin{array}{ccc}
\lambda_{{(s)}\dot{\alpha} \dot{a}}  & 0 & 0\\
0 & \lambda_{{(g)}\dot{\alpha} \dot{a}} & 0\\
0 & 0 & {-\lambda_{{(g)}\dot{\alpha} \dot{a}}}^T
\end{array}\right)\\
& M^{\alpha3}=\left(\begin{array}{ccc}
0 & M_{(gs)}^{\alpha} & 0\\
0 & 0 & M_{(g)}^{\alpha}\\
{M_{(gs)}^{\alpha}}^T & 0 & 0
\end{array}\right)
&&\lambda_{\dot{\alpha}3}=\left(\begin{array}{ccc}
0 & 0 & \lambda_{{(gs)}\dot{\alpha}}\\
{-\lambda_{{(gs)}\dot{\alpha}}}^T & 0 & 0\\
0 & \lambda_{{(g)}\dot{\alpha}} & 0
\end{array}\right)\\
& M^{\alpha4}=\left(\begin{array}{ccc}
0 & 0 & M_{(gs)}^{'\alpha}\\
{M_{(gs)}^{'\alpha}}^T & 0 & 0\\
0 & M_{(g)}^{'\alpha} & 0
\end{array}\right)
&&\lambda_{\dot{\alpha}4}=\left(\begin{array}{ccc}
0 & \lambda'_{{(gs)}\dot{\alpha}} & 0\\
0 & 0 & \lambda'_{{(g)}\dot{\alpha}}\\
{-\lambda'_{{(gs)}\dot{\alpha}}}^T & 0 & 0
\end{array}\right)
\end{align}~~~~~~~
\end{subequations} 
where $\alpha, \dot\alpha, \dot{a}=1,2$ and 
\begin{subequations}
\begin{align}
&M_{(s)}^{\alpha \dot{a}}={M_{(s)}^{\alpha \dot{a}}}^T
&&\lambda_{{(s)}\dot{\alpha} \dot{a}}={-\lambda_{{(s)}\dot{\alpha} \dot{a}}}^T\\
&M_{(g)}^{\alpha}={M_{(g)}^{\alpha}}^T
&&\lambda_{{(g)}\dot{\alpha}}={-\lambda_{{(g)}\dot{\alpha}}}^T\\
&M_{(g)}^{'\alpha}={M_{(g)}^{'\alpha}}^T
&&\lambda'_{{(g)}\dot{\alpha}}={-\lambda'_{{(g)}\dot{\alpha}}}^T.
\end{align}~~~~~~~~~
\end{subequations}
\subsection{Charged Bosonic Moduli}\label{chargedm}

The charged bosonic moduli are denoted by $w_{\dot\alpha}$ and $\bar w_{\dot\alpha}$ and are basically the NS sector of the string states 
stretching between 
D3-branes and D-instantons. These moduli experience two different boundary conditions at the string endpoints. The 
NS sector with such mixed boundary conditions is a Weyl spinor state in the Lorentz space and a boson from the internal space point of view. 
It is then invariant under the orbifold transformations. The charged bosonic moduli are ruled out in the spectrum of 
pure stringy set-up discussed in detail in \cite{Ghorbani:2010ks}. They are however present in the gauge instanton configuration and play the role of the
size of the instanton. 
As mentioned in the paragraph before equation (\ref{achi}) the orientifold action on Chan-Paton factors with 
at least one endpoint on D3-brane is given by an antisymmetric matrix denoted by $\gamma_-(\Omega)$:
\begin{equation}\label{gamma-}
 \gamma_{-}\left(\Omega\right)=\left(\begin{array}{ccc}
\epsilon_{N_{1}\times N_{1}} & 0 & 0\\
0 & 0 & \one_{N\times N}\\
0 & -\one_{N\times N} & 0
\end{array}\right)
\end{equation}
where $\epsilon$ is an antisymmetric matrix with $\epsilon^2=-\one$. The matrix $\epsilon$ however do not enter in the calculations
because in our model we do not have any gauge branes on node 1. 
The modulus $w_{\dot\alpha}$ ($\bar w_{\dot\alpha}$) is in the fundamental (anti-fundamental) representation of gauge instanton group, i.e. U$(k_g)$
and in anti-fundamental (fundamental) representations of the gauge group U$(N)$. Therefore the Chan-Paton matrices should satisfy the following conditions:
\begin{equation}
  \bar w_{\dot{\alpha}}=\gamma\left(g\right)w_{\dot{\alpha}}\gamma\left(g\right)^{-1},~~~~~~~
\bar w_{\dot\alpha}=\gamma_{+}\left(\Omega\right) w_{\dot\alpha}^T \gamma_{-}\left(\Omega\right)^{-1}\\
\end{equation}
where
\begin{equation}
 w_{\dot{\alpha}}=\left(\begin{array}{ccc}
0 & 0 & 0\\
0 & w_{{(g)}\dot{\alpha}} & 0\\
0 & 0 & w'_{{(g)}\dot{\alpha}}
\end{array}\right),~~~~~
\bar{w}_{\dot{\alpha}}=\left(\begin{array}{ccc}
0 & 0 & 0\\
0 & {w'_{{(g)}\dot{\alpha}}}^T & 0\\
0 & 0 & {-w_{{(g)}\dot{\alpha}}}^T
\end{array}\right).
\end{equation}

There are two independent entries $w_{{(g)}\dot{\alpha}}$ and $w'_{{(g)}\dot{\alpha}}$ in the Chan-Paton matrix $w_{\dot{\alpha}}$
which are both $k_g\times N$ matrices. The entries in matrix $\bar{w}_{\dot{\alpha}}$ are $N\times k_g$ matrices which are 
related to $w_{{(g)}\dot{\alpha}}$ and $w'_{{(g)}\dot{\alpha}}$. The subscript ``g" again emphasizes the gauge characteristics of these moduli.
\subsection{Charged Fermionic Moduli}

The charged fermionic moduli are denoted by $\mu^A$ and $\bar\mu^A$ with $A=1,...,4$ and stem from the R sector of the 
open string state with mixed Neumann-Dirichlet 
boundary condition at the string endpoints.  The orbifold and orientifold backgrounds impose the following conditions on the Chan-Paton factors:
\begin{equation}
  \bar\mu^{A}=R(g)^A_{~B} \gamma(g) \mu^{B} \gamma(g)^{-1}~~~~~~\bar\mu^{A}=R(\Omega)^A_{~B} \gamma_+(\Omega) (\mu^{B})^T \gamma_-(\Omega)^{-1}
\end{equation}
The Chan-Paton matrices which survive the above conditions are:
\begin{subequations}
\begin{align}
& \mu^{\dot a}=\left(\begin{array}{ccc}
0 & 0 & 0\\
0 & \mu^{a}_{(g)} & 0\\
0 & 0 & \mu'^{a}_{(g)}
\end{array}\right)~~~~~
&&\bar{\mu}^{\dot a}=\left(\begin{array}{ccc}
0 & 0 & 0\\
0 & {\mu'^{a}_{(g)}}^T & 0\\
0 & 0 & {-\mu^{a}_{(g)}}^T
\end{array}\right)
\end{align}
\begin{align}
& \mu^{3}=\left(\begin{array}{ccc}
0 & 0 & 0\\
0 & 0 & \mu_{(g)}\\
\mu_{(s)} & 0 & 0
\end{array}\right)~~~~~
&&\bar{\mu}^{3}=\left(\begin{array}{ccc}
0 & {\mu_{(s)}}^T & 0\\
0 & 0 & {-\mu_{(g)}}^T\\
0 & 0 & 0
\end{array}\right)
\end{align}
\begin{align}
& \mu^{4}=\left(\begin{array}{ccc}
0 & 0 & 0\\
\mu'_{(s)} & 0 & 0\\
0 & \mu'_{(g)} & 0
\end{array}\right)~~~~~
&&\bar{\mu}^{4}=\left(\begin{array}{ccc}
0 & 0 & {-\mu'_{(s)}}^T\\
0 & 0 & 0\\
0 & {\mu'_{(g)}}^T & 0
\end{array}\right).
\end{align}
\end{subequations}\\
where $\dot a=1,2$ denotes the index of SU$(2)_L$ in the internal space. The moduli in this sector are
either of stringy or gauge type. Despite the charged bosonic moduli, 
the Chan-Paton matrices in the charged fermionic sector can be both diagonal and off-diagonal. 
For generic $k_g$ ($k_s$) the entries of $\mu^{\dot a}$ and $\bar\mu^{\dot a}$ with $\dot a=1,2$ are respectively $k_g \times N$ ($k_s \times N$) 
and $N \times k_g$ ($N \times k_s$) matrices. 
\subsection{Auxiliary Moduli}

In order to reduce the quartic terms in the action into quadratic terms and construct the BRST pairs we define some auxiliary moduli. Their 
Chan-Paton factors are determined by means of their definitions in terms of the moduli. For quartic terms in $a^\mu$ and $\chi^m$ with 
$m=4,5,6,7$ we define:
\begin{equation}
 D^c\equiv \bar{\eta}_{\mu\nu}^{c}\left[a^{\mu},a^{\nu}\right]+\bar{\zeta}_{mn}^{c}\left[\chi^{m},\chi^{n}\right].
\end{equation}
We will address the tensors $\bar{\eta}_{\mu\nu}^{c}$ and $\bar{\zeta}_{mn}^{c}$ in appendix (\ref{action}). The Chan-Paton matrix for $D^c$
reads:
\begin{equation}
D^c=\left(\begin{array}{ccc}
D_{(s)}^{c} & 0 & 0\\
0 & D_{(g)}^{c} & 0\\
0 & 0 & {-D_{(g)}^{c}}^T
\end{array}\right).
\end{equation}
where $D_{(s)}^c=-{D^c_{(s)}}^T$. The auxiliary moduli that pairs up with the off-diagonal moduli $M^{\alpha \dot a}$ is defined as:
\begin{equation}
 C^{\alpha a}\equiv (\bar{\sigma}_{\mu})^{~\alpha}_{\dot{\alpha}}
 \left[a^{\mu},\chi^{\dot\alpha a}\right]
\end{equation}
where $a=3,4$ and $(\bar{\sigma}_{\mu})^{~\alpha}_{\dot{\alpha}}$
has been defined in appendix A in \cite{Billo:2002hm}. The Chan-Paton matrices for $C^{\alpha a}$ read:
\begin{equation}
C^{\alpha 3}=\left(\begin{array}{ccc}
0 & C_{(gs)}^{\alpha} & 0\\
0 & 0 & C_{(g)}^{\alpha}\\
{C_{(gs)}^{\alpha}}^T & 0 & 0
\end{array}\right)~~~~~~~~~
C^{\alpha 4}=\left(\begin{array}{ccc}
0 & 0 & C_{(gs)}^{'\alpha}\\
{C_{(gs)}^{'\alpha}}^T & 0 & 0\\
0 & C_{(g)}^{'\alpha} & 0
\end{array}\right)
\end{equation}
with 
\begin{equation}
C_{(g)}^{\alpha}={C_{(g)}^{\alpha}}^T ~~~~~~~~~~~~~~~~~~~~C_{(g)}^{'\alpha}={C_{(g)}^{'\alpha}}^T
\end{equation}
where $\alpha=1,2$ stands for SU$(2)_L$ subgroup of the Lorentz space. As seen from the subscripts of the entries there are gauge and 
gauge-stringy types for this auxiliary modulus.\\
The auxiliary moduli which pair up with charged moduli $\mu^{a}$ is
\begin{equation}
h_{a} \equiv \bar{w}^{\dot{\alpha}} \chi_{\dot\alpha a}
\end{equation}
The corresponding Chan-Paton matrices takes the same structure as moduli $\mu^a$:
\begin{subequations}
\begin{align}
&h^{3}=\left(\begin{array}{ccc}
0 & 0 & 0\\
0 & 0 & h_{(g)}\\
h_{(s)} & 0 & 0
\end{array}\right)
&&\bar{h}^{3}=\left(\begin{array}{ccc}
0 & h_{(s)}^T & 0\\
0 & 0 & {-h_{(g)}}^T\\
0 & 0 & 0
\end{array}\right)
\end{align}
\begin{align}
&h^{4}=\left(\begin{array}{ccc}
0 & 0 & 0\\
h'_{(s)} & 0 & 0\\
0 & h'_{(g)} & 0
\end{array}\right)
&&\bar{h}^{4}=\left(\begin{array}{ccc}
0 & 0 & -{h'_{(s)}}^T\\
0 & 0 & 0\\
0 & {h'_{(g)}}^T & 0
\end{array}\right).
\end{align}
\end{subequations}
\section{Gauge-Stringy Moduli Action}\label{sec-action}
The instanton moduli action is known from the ADHM instanton construction. It can also be derived from string theory framework 
\cite{Billo:2002hm,Dorey:2002ik} by calculating the various disk amplitudes to find all possible couplings. In appendix
\ref{action} the general action of the moduli in D3/D(--1) system has been given. The full action for such a set-up is
described by fields with indices being 
either along the D3-brane world volume (Lorentz space) or orthogonal to it (internal space). In our particular model due to the
presence of the orbifold background $\mathbb{Z}_3$ acting on the first two complex internal directions we need further split of
internal coordinates in the action. This has been done as well in appendix \ref{action}. 

The generic moduli action after applying the localization (see \ref{local}) reads:
 \begin{equation}\label{totaction}
 \begin{split}
S_{\Omega}=\frac{1}{2}t{D}_{c}^{2}
-\frac{1}{2}tC^{\alpha a}C_{\alpha a}-it{\lambda}_{c}Q_\Omega^{2}\left({\lambda}^{c}\right)
-\frac{i}{2}t{M}^{\alpha a}Q_\Omega^{2}\left({M}_{\alpha a}\right)\\
 -2it{\bar{h}}^{a}{h}_{a}+2it{\bar{\mu}}^{a}Q_\Omega^{2}\left({\mu}_{a}\right)
 +u\bar{f}_{\mu\nu}{M}^{\mu}{M}^{\nu}
 +u\bar{f}_{\mu\nu}{a}^{\mu}Q_\Omega^{2}\left({a}^{\nu}\right)\\
 +u\bar{f}_{\mu\nu}\left(\bar{\sigma}^{\mu\nu}\right)_{\dot{\alpha}\dot{\beta}}{\bar{\mu}}^{\dot{\alpha}}{\mu}^{\dot{\beta}}
 +u\bar{f}_{\mu\nu}\left(\bar{\sigma}^{\mu\nu}\right)_{\dot{\alpha}\dot{\beta}}{\bar{w}}^{\dot{\alpha}}Q_\Omega^{2}\left({w}^{\dot{\beta}}\right)\\
 +u\bar{f}_{\mu\nu}\left(\bar{\sigma}^{\mu\nu}\right)_{\,\,\dot{\beta}}^{\dot{\alpha}}{\lambda}_{\dot{\alpha}a}{\lambda}^{\dot{\beta}a}
 +u\bar{f}_{\mu\nu}\left(\bar{\sigma}^{\mu\nu}\right)_{\,\,\dot{\beta}}^{\dot{\alpha}}{\chi}_{\dot{\alpha}a}Q_\Omega^{2}\left({\chi}^{\dot{\beta}a}\right)
   \end{split}
\end{equation}
where $t=x^4/g_{0}^{2}$ and $u=z/x^2 g_{0}^{2}$. 
The subscript $\Omega$ denotes the presence of the $\Omega$-background that in our set-up is the Ramond-Ramond 3-form.
The BRST-charge $Q_\Omega$ transformation is given in (\ref{QOmega}). Note that in eq. (\ref{totaction}) the moduli are $3\times 3$ block matrices.
To calculate the partition function one has to expand the action in terms of the Chan-Paton entries which are gauge, 
stringy or gauge-stringy moduli. The action can then be decomposed into three parts:

\subsection{Gauge Moduli Action}
 \begin{equation}\label{Sg}
 \begin{split}
S_{\Omega}^{(g)}=tD_{(g)}^{c}D_{(g)c}-2it{\lambda_{(g)}}_{c}Q_{\Omega}^{2}\left({\lambda_{(g)}}^{c}\right)\\
-\frac{1}{2}tC_{(g)}^{\alpha}C'_{(g)\alpha}-it{M}_{(g)}^{\alpha}Q_{\Omega}^{2}\left({M'}_{(g)\alpha}\right)\\
 +4it\left(h_{(g)}^{T}h'_{(g)}+h_{(g)}'h{}_{(g)}^{T}\right)-2it\left[\mu_{(g)}^{T}Q_{\Omega}^{2}\left(\mu'_{(g)}\right)+\mu_{(g)}'^{T}Q_{\Omega}^{2}\left(\mu{}_{(g)}\right)\right]\\
 +u\bar{f}_{\mu\nu}{M_{(g)}}^{\mu}{M_{(g)}}^{\nu}+u\bar{f}_{\mu\nu}{a_{(g)}}^{\mu}Q_{\Omega}^{2}\left({a_{(g)}}^{\nu}\right)\\
 -u\bar{f}_{\mu\nu}\left(\bar{\sigma}^{\mu\nu}\right)_{\dot{\alpha}\dot{\beta}}\mu{}_{(g)}^{\dot{\alpha}}{}^{T}\mu_{(g)}'^{\dot{\beta}}-u\bar{f}_{\mu\nu}\left(\bar{\sigma}^{\mu\nu}\right)_{\dot{\alpha}\dot{\beta}}w_{(g)}^{\dot{\alpha}T}Q_{\Omega}^{2}\left(w_{(g)}'^{\dot{\beta}}\right)\\
 +u\bar{f}_{\mu\nu}\left(\bar{\sigma}^{\mu\nu}\right)_{\,\,\dot{\beta}}^{\dot{\alpha}}\lambda_{(g)\dot{\alpha}}\lambda_{(g)}'^{\dot{\beta}}+u\bar{f}_{\mu\nu}\left(\bar{\sigma}^{\mu\nu}\right)_{\,\,\dot{\beta}}^{\dot{\alpha}}\chi_{(g)\dot{\alpha}}\chi_{(g)}'^{\dot{\beta}}\\
  \end{split}
\end{equation}
This action is obtained by setting to zero all the entries of the moduli Chan-Paton matrices except those with subscript ''g``.  	
The BRST $Q_\Omega$ transformation is still given by eq. (\ref{QOmega}) replacing only the $3$ by $3$ Chan-Paton matrices by the gauge moduli entries. 

\subsection{Stringy Moduli Action}
 \begin{equation}\label{Ss}
 \begin{split}
S_{\Omega}^{(s)}=\frac{1}{2}t{D}_{(s)}^{c}{D}_{(s)c}-it{\lambda}_{(s)c}Q_{\Omega}^{2}\left({\lambda}_{(s)}^{c}\right)\\
 -4it h_{(s)}^{T}h'_{(s)}+4it\mu_{(s)}^{T}Q_{\Omega}^{2}\left(\mu'_{(s)}\right)\\
 +u\bar{f}_{\mu\nu}{M_{(s)}}^{\mu}{M_{(s)}}^{\nu}+u\bar{f}_{\mu\nu}{a_{(s)}}^{\mu}Q_{\Omega}^{2}\left({a_{(s)}}^{\nu}\right)\\
  \end{split}
\end{equation}
This action is similarly obtained by keeping only the stringy moduli entries in the Chan-Paton matrices. The BRST charge $Q_\Omega$ acts as well
only on stringy matrices. This result was obtained and used in stringy multi-instanton calculus in \cite{Ghorbani:2010ks}.

\subsection{Gauge-Stringy Moduli Action}
\begin{equation}\label{Sgs}
 \begin{split}
S_{\Omega}^{(gs)}= S_\Omega^{(g)}+ S_\Omega^{(s)}-t\left[C{}_{(gs)\alpha}C_{(gs)}^{\alpha'T}+C_{(gs)}^{\alpha T}C'_{(gs)\alpha}\right]\\
 -is\left[{M}_{(gs)}^{\alpha}Q_{\Omega}^{2}\left({M'}_{(gs)\alpha}^{T}\right)+{M}_{(gs)}^{\alpha T}Q_{\Omega}^{2}\left({M'}_{(gs)\alpha}\right)\right]\\
 -u\bar{f}_{\mu\nu}\left(\bar{\sigma}^{\mu\nu}\right)_{\,\,\dot{\beta}}^{\dot{\alpha}}\left(\lambda_{(gs)\dot{\alpha}}^{T}\lambda_{(gs)}'^{\dot{\beta}}+\lambda_{(gs)\dot{\alpha}}'^{T}\lambda_{(gs)}^{\dot{\beta}}\right)\\
 -u\bar{f}_{\mu\nu}\left(\bar{\sigma}^{\mu\nu}\right)_{\,\,\dot{\beta}}^{\dot{\alpha}}\left(\chi_{(gs)\dot{\alpha}}^{T}Q_{\Omega}^{2}\chi_{(gs)}'^{\dot{\beta}}+\chi_{(gs)\dot{\alpha}}'^{T}Q_{\Omega}^{2}\chi_{(gs)}^{\dot{\beta}}\right)\\
   \end{split}
\end{equation}
The gauge-stringy action is actually obtained by keeping all the entries in the Chan-Paton matrices. As we see this is not only the 
sum of gauge and stringy actions but it contains new interactions due to the extra gauge-stringy zero modes that play a 
crucial r\^ole in gauge-stringy instanton calculus. 

\section{Dimensionless Moduli Measure}\label{sec-dimless}

The moduli action should be a dimensionless quantity. Each modulus in the action then acquires a canonical dimension in general.
The canonical length dimensions of the moduli appeared in the action and that of the auxiliary moduli are demonstrated in table 2~. 
Recall that the Chan-Paton factors are $3\times 3$ block matrices whose entries are the gauge, stringy or gauge-stringy 
moduli matrices. In order to count the number of degrees of freedom for Chan-Paton entries one should be careful about the 
associate group representation each entry belongs to. 
In table 2 considering this point the number of degrees of freedom for different types of moduli is presented. Note that 
in this table we have not considered the number of degrees of freedom coming from the moduli indices. However to obtain the dimension of the 
measure of the path integral one should keep in mind to consider the indices too, as well as different length dimensions in each BRST pair. 

To see how instantons contribute in the low-energy effective action one needs to integrate out all moduli present in the theory. 
Through the interactions of the gauge sector of the supersymmetry theory with the changed moduli (and therefore indirectly interacting with 
neutral moduli) the instantonic effects in the effective action of the theory (here the prepotential of $\mathcal N=2$) can be extracted. 
One therefore deals with the evaluation of the instanton partition function in the following general form:
\begin{equation}\label{Zkgks}
Z_{k_g,k_s}=\mu^{\gamma(k_g,k_s)} e^{-S_{cl}} \int {d\mathcal M ~ e^{-S_{\Omega}}}
\end{equation}
where $d\mathcal M$ denotes the moduli measure of integral which contains the eight superspace coordinates $x^\mu\equiv \text{diag}(a^\mu)$
and $\theta^{\alpha \dot a}=\text{diag}(M^{\alpha \dot a})$. $S_{cl}=-8\pi^2 k/g_s$ is the instanton classical action with 
$g_s=g_{YM}^2$ being the string coupling constant. $S_{\Omega}$ is either the gauge, stringy or gauge-stringy moduli action in the presence of the $\Omega$-background, given 
in equations (\ref{Ss})-(\ref{Sgs}) for the theory we discuss in this article. The instanton moduli measure is usually dimensionful. The factor  
$\mu^{\gamma(k_g,k_s)}$ in front of the integral is a dimensionful parameter in order to compensate the dimension of the moduli measure $d\mathcal M$
to have in the end of the day a dimensionless partition function $Z_{k_g,k_s}$.

In the most general case $Z_{k_g,k_s}$ is the partition function of the gauge-stringy instanton.
One can obtain the gauge (stringy) partition functions by setting $k_s=0$ ($k_g=0$). This will discard 
all the stringy (gauge) moduli as well as the gauge-stringy zero modes from the moduli action. 

From table 2 it is easily seen that the appropriate dimensionful factor in (\ref{Zkgks}) to compensate the dimension of the 
total moduli measure $d\mathcal M$ is:
\begin{equation}\label{measuredim}
 \mu^\gamma=\mu^{b_1(k_g-k_s)}
\end{equation}
where $b_1=N-2$ is the one-loop coefficient of the $\beta$-function for $\mathcal N=2$ SU$(N)$ SYM theory with one matter multiplet
in symmetric representation. 

To get (\ref{measuredim}) one should note that the dimension of the
differential of a fermionic modulus, $dF$, is the inverse of the dimension of the fermionic modulus $F$. An interesting observation is that 
despite introducing four new gauge-stringy zero modes $(M^{\alpha a},C^{\alpha a})$ and $(\chi_{\dot\alpha a},\lambda_{\dot\alpha a})$ 
their dimensions in the measure cancel out among themselves so that the dimension of the gauge-stringy measure is simply the sum 
of the dimensions of the gauge and stringy moduli measures. Let us investigate the situation that $k_g$ and $k_s$ take 
different values in (\ref{measuredim}).

If we put $k_s=0$ we are left with a pure gauge instanton partition function. In this case $S_\Omega$ is the gauge 
instanton moduli action and the dimensionful parameter $\mu^\gamma=\mu^{-k_gb_1}$ corresponds to the Pauli-Villars 
renormalization mass scale. The combination of this factor and classical
instanton action describes the renormalization group invariant $\Lambda$-parameter (see e.g. \cite{Bianchi:2007ft} and \cite{Ghorbani:2011xh}).

For the case of the pure stringy instanton set-up i.e. when $k_g=0$ the interpretation in the above paragraph is no more valid. 
The dimensionful parameter in (\ref{Zkgks}) is given now by $\mu^\gamma=\mu^{-k_sb_1}$, where $\mu$ 
is proportional to $1/\sqrt{\alpha'}$, the string tension. The stringy corrections to
the low-energy effective action carry therefore the string theory parameter $\alpha '$. 
Any choice $k_g>k_s$ ($k_g<k_s$) is equivalent to
gauge (stringy) instanton cases. 

The most interesting choice takes place when $k_g=k_s$. In this case that the number of D-instantons
on nodes 1 and 2 (3) is equal, the total moduli measure becomes dimensionless.
The prepotential corrections due to various instanton charges should be considered on the same foot; they 
contribute equally in the effective action. The string theory parameter $\alpha'$ do not enter in the prepotential corrections.
Having a dimensionless measure happens alternatively when $b_1=0$, i.e. where 
the theory is superconformal. An example of such a model was introduced in \cite{Ghorbani:2010ks}. 
There the stringy corrections to the prepotential in $\mathcal N=2$ SU$(2)$ with symmetric matter (which turns out to 
be a superconformal theory) was calculated.

\begin{table}[h]\footnotesize
\begin{center}
\begin{tabular}{|c|c|c|c|c|}
\hline 
 $\downarrow$ BRST pairs $|$ moduli $\rightarrow$ & gauge  & stringy & gauge-stringy  & $\left[L\right]$  \tabularnewline
\hline 
\hline 
$\left(a^{\mu},M^{\mu}\right)$ & $k_{g}^{2}$ & $\frac{1}{2}k_{s}\left(k_{s}+1\right)$ & $\times$ & $(L,L^{\frac{1}{2}})$  \tabularnewline
\hline 
$\left(\bar{\chi},\eta\right)$ & $k_{g}^{2}$ & $\frac{1}{2}k_{s}\left(k_{s}-1\right)$ & $\times$ & $(L^{-1},L^{-\frac{3}{2}})$  \tabularnewline
\hline 
$\left(\lambda^{c},D^{c}\right)$ & $k_{g}^{2}$ & $\frac{1}{2}k_{s}\left(k_{s}-1\right)$ & $\times$ & $(L^{-\frac{3}{2}},L^{-2})$  \tabularnewline
\hline 
$\left(\mu^{a},h^{a}\right),\left(\bar{\mu}^{a},\bar{h}^{a}\right)$ & $k_{g}N$ & $k_{s}N$ & $\times$ & $(L^{\frac{1}{2}},L^{0})$ \tabularnewline
\hline 
$\left(w^{\dot{\alpha}},\mu^{\dot{\alpha}}\right),\left(\bar{w}_{\dot{\alpha}},\bar{\mu}_{\dot{\alpha}}\right)$ & $k_{g}N$ & $\times$ & $\times$ & $(L,L^{\frac{1}{2}})$  \tabularnewline
\hline 
$\left(M^{\alpha a},C^{\alpha a}\right)$ & $\frac{1}{2}k_{g}\left(k_{g}+1\right)$ & $\times$ & $k_{g}k_{s}$ & $(L^{\frac{1}{2}},L^{0})$  \tabularnewline
\hline 
$\left(\chi_{\dot{\alpha}a},\lambda_{\dot{\alpha a}}\right)$ & $\frac{1}{2}k_{g}\left(k_{g}-1\right)$ & $\times$ & $k_{g}k_{s}$ & $(L^{-1},L^{-\frac{3}{2}})$  \tabularnewline
\hline 
\end{tabular}
 \caption{Number of degrees of freedom for gauge, stringy and gauge-stringy BRST pairs and the canonical 
 length dimension for each pair. For a given BRST pair $(\Psi_0,\Psi_1)$ where $\Psi_1=Q\Psi_0$ the length dimension is given by 
 $(L^\Delta, L^{\Delta-\frac{1}{2}})$ with $\Delta$ being the length dimension of $\Psi_0$.}
\end{center}
 \end{table}
 

 \section{Partition Functions}\label{sec-patition}
 The total partition function is obtained by summing over all $k$-instanton partition functions. For our gauge-stringy model 
 the total partition function reads:
 \begin{equation}
  Z=\sum_{k_g,k_s=1}^{\infty}{\mu^{b_1 (k_g-k_s)}e^{2\pi i\tau(k_g,k_s)} Z_{k_g,k_s}}
 \end{equation}
where $b_1=N-2$ and $\tau$ is the complex coupling constant depending on both gauge and stringy instanton charges, $k_g$ and $k_s$. 
We will assume in all our calculations the special case $k_g=k_s\equiv k$, 
the partition function then simplifies to 
\begin{equation}
   Z=\sum_{k=1}^{\infty}{e^{2\pi i\tau} Z_k},
\end{equation}
where $k$ is the gauge-stringy winding number and $Z_k$ is the dimensionless $k$-instanton partition function. 
From (\ref{Zkgks}) we have
\begin{equation}\label{Z_kgs}
 Z_k^{(gs)}= \int d\mathcal{ M}_{(g)} d\mathcal{ M}_{(s)} 
 d\mathcal{ M}_{(gs)} e^{-S_{(g)}-S_{(s)}-S_{(gs)}}
\end{equation}
where $S_{(g)}, S_{(s)}$ and $S_{(gs)}$ are given in (\ref{Sg})-(\ref{Sgs}) and 
\begin{subequations}
\begin{align}
\begin{split}
 d\mathcal{ M}_{(g)}=d\chi_{(g)}da^\mu_{(g)}dM^\mu_{(g)}dD^c_{(g)}d\lambda^c_{(g)}d\mu_{(g)}
 d\mu'_{(g)} dh_{(g)} dh'_{(g)} d\chi^{\dot\alpha}_{(g)}  d\chi'^{\dot\alpha}_{(g)}\\
\times d\lambda^{\dot\alpha}_{(g)} d\lambda'^{\dot\alpha}_{(g)}
dw^{\dot\alpha}_{(g)} dw'^{\dot\alpha}_{(g)} d\mu^{\dot\alpha}_{(g)} d\mu'^{\dot\alpha}_{(g)}
 dC^{\alpha}_{(g)} dC'^{\alpha}_{(g)} dM^{\alpha}_{(g)} dM'^{\alpha}_{(g)}
\end{split}
\end{align}
\begin{align}
 d\mathcal{ M}_{(s)}=d\chi_{(s)}da^\mu_{(s)}dM^\mu_{(s)}dD^c_{(s)}d\lambda^c_{(s)}d\mu_{(s)}
 d\mu'_{(s)} dh_{(s)} dh'_{(s)} 
\end{align}
\begin{align}
 d\mathcal{M}_{(gs)}=d\chi_{(g)}d\chi_{(s)}dC^{\alpha}_{(gs)} dC'^{\alpha}_{(gs)} dM^{\alpha}_{(gs)} dM'^{\alpha}_{(gs)}.
 \end{align}
\end{subequations}
The primed and unprimed moduli have been introduced in the entries of the Chan-Paton matrices in section \ref{sec-spectrum}.
The integration over all moduli except the unpaired moduli $\chi_{(g)}$ and $\chi_{(s)}$ can be performed easily. 
The integration over paired moduli is of Gaussian type so the result is the determinant of the BRST charge in the appropriate
symmetry group representation. The results of the integral (\ref{Z_kgs}) can be summarized as follows:

\begin{equation}\label{Ztot}
 Z_k^{(gs)}=\int{d\chi d\tilde\chi}~ \mathcal{I}_{(g)} ~\mathcal{I}_{(s)} ~ \mathcal{I}_{(gs)}
\end{equation}
where
\begin{equation}\label{Ig}
\mathcal{I}_{(g)}= \frac{\mathcal P_{(g)}(\tilde\chi) \mathcal R_{(g)}(\tilde\chi) \mathcal C_{(g)}(\tilde\chi)}
{\mathcal Q_{(g)}(\tilde\chi) \mathcal L_{(g)}(\tilde\chi) \mathcal W_{(g)}(\tilde\chi) } 
\end{equation}
\begin{equation}\label{Is}
\mathcal{I}_{(s)}= \frac{\mathcal P_{(s)}(\chi) \mathcal R_{(s)}(\chi)}
{\mathcal Q_{(s)}(\chi)} 
\end{equation}
\begin{equation}\label{Igs}
\mathcal{I}_{(gs)}= \frac{ \mathcal C_{(gs)}(\tilde\chi,\chi)}
{ \mathcal L_{(gs)}(\tilde\chi,\chi) } 
\end{equation}
For simplicity we relabel the gauge instanton parameter as $\tilde\chi\equiv \chi_{(g)}$.
The functions in the numerator and denominator of eqs. (\ref{Ig}) and (\ref{Is}) are defined in the same way for gauge and stringy subscripts are:
\begin{equation}\label{QgQs}
\begin{split}
\mathcal Q_{(g)}(\tilde\chi),Q_{(s)}(\chi) 
& \equiv \int da_{\mu}dM_{\mu} \\
& \times \exp\left\{ u\bar{f}_{\mu\nu}[a_{\mu}Q_\Omega^{2}\left(a_{\nu}\right)+M_{\mu}M_{\nu}]\right\} 
=\text{det}_{(a^\mu,M^\mu)}{(Q_\Omega^2)}^{-1/2}
\end{split}
\end{equation}
\begin{equation}\label{PgPs}
\begin{split}
\mathcal P_{(g)}(\tilde\chi),\mathcal P_{(s)}(\chi)
&\equiv \int dD_{{c}}d\lambda_{{c}} \\
& \times \exp{\{-\frac{t}{2}[\lambda_{{c}}Q_\Omega^{2}
 \left(\lambda^{{c}}\right)+D_{{c}}D^{{c}}]} \}
 =\text{Pf}_{(\lambda^c,D^c)}{(Q_\Omega^2)}
\end{split}
\end{equation}
 \begin{equation}\label{RgRs}
\begin{split}
 \mathcal R_{(s)}(\tilde\chi),\mathcal R_{(s)}(\chi) 
 &\equiv \int d\mu dh d\mu'dh' \\
 &\times \exp\left\{ -th^{T}h'+t\mu^{T}Q_\Omega^{2}\left(\mu'\right)\right\}
=\text{det}_{(\mu,h)}{(Q_\Omega^2)}
\end{split}
\end{equation}
The gauge or stringy subscripts is understood in the right hand side of equations (\ref{QgQs})-(\ref{RgRs}). 
Note that the functions $\mathcal P,\mathcal Q$ and $\mathcal R$ for gauge instanton are functions of modulus 
$\tilde\chi\equiv\chi_{(g)}$ while those of stringy type are functions of $\chi\equiv \chi_{(s)}$. 
Consequently the result of the above integrations, say function $\mathcal P$, is different for gauge and stringy 
functions and depends on the group representation the moduli in the function belong to. The functions which are defined similarly for 
gauge and gauge-stringy instantons are as the following:
\begin{equation}
\begin{split}
 \mathcal L_{(g)}(\tilde\chi),\mathcal L_{(gs)}(\tilde\chi,\chi)
& \equiv \int  d{\chi}_{\dot{\alpha}}d{\lambda}^{\dot{\alpha}}
 d{\chi}'_{\dot{\alpha}} d{\lambda}'^{\dot{\alpha}} \\
& \times\exp\{ u\bar{f}^{\mu\nu}(\bar{\sigma}_{\mu\nu})^{\dot{\alpha}\dot{\beta}}[{\lambda}_{\dot{\alpha}}^{T}
 {\lambda}'_{\dot{\beta}}+
 {\chi}_{\dot{\alpha}}^{T}Q_\Omega^{2}({\chi}_{\dot{\beta}}')]\} 
=\text{det}_{(\chi^{\dot\alpha},\lambda^{\dot\alpha})}{(Q_\Omega^2)}^{-1}
 \end{split}
\end{equation}
 \begin{equation}
\begin{split}
 \mathcal{C}_{(g)}(\tilde\chi),\mathcal{C}_{(gs)}(\tilde\chi,\chi)
 &\equiv \int d{C}_{\alpha}d{M}^{\alpha} d{C}_{\alpha}'d{M}'^{\alpha}\\
&\times \exp\{-\frac{t}{2}[{M}^{\alpha T}Q_\Omega^{2}({M}_{\alpha}')+{C}_{\alpha}^{T}{C}'^{\alpha}]\} 
 =\text{det}_{(M^\alpha,C^\alpha)}{(Q_\Omega^2)}
\end{split}
\end{equation}
For gauge instantons, the above functions depend only on $\tilde\chi$ and for gauge-stringy instantons they are functions of 
both $\tilde\chi$ and $\chi$. The function which appears only in the integrand of gauge instanton partition function is
\begin{equation}\label{Wg}
\begin{split}
 \mathcal W_{(g)}(\tilde\chi) 
&\equiv\int d{w}^{\dot{\alpha}}d{\mu}^{\dot{\alpha}}
  d{w}'^{\dot{\alpha}}d{\mu}'^{\dot{\alpha}}\\
 & \times \exp\left\{ u\bar{f}_{\mu\nu}\left(\bar{\sigma}^{\mu\nu}\right)_{\dot{\alpha}\dot{\beta}}\left[
  {\mu}^{\dot{\alpha}T}{\mu}'^{\dot{\beta}}+{w}^{\dot{\alpha}T}Q_\Omega^{2}
  \left({w}'^{\dot{\alpha}}\right)\right]\right\} 
   =\text{det}_{(w^{\dot\alpha},\mu^{\dot\alpha})}{(Q_\Omega^2)}^{-1}.
\end{split}
\end{equation}
To perform the above integrations one needs to know the transformation properties of each BRST pair. 
The determinant of the BRST operators may be interpreted as the product of all non-zero eigenvalues in a certain 
symmetry group representation. One therefore needs to know the weight vectors of all symmetry groups in the theory. 
In table 2 the symmetry group and the representation of each pair has been demonstrated. 

\begin{table}
\begin{center}
 \begin{tabular}{|c|c|c|c|c|}
\hline 
$(\Psi_{0},\Psi_{1})$ & SO$(k_{s})$ & U$(k_{g})$ & SU$(N)$ & SU$(2)\times$SU$(2)'$\tabularnewline
\hline 
\hline 
$(a_{(s)}^{\mu},M_{(s)}^{\mu})$ &  \tiny\Yvcentermath1$\yng(2)$ & $\bullet$ & $\bullet$ & $(2,2)$\tabularnewline
\hline 
$(D_{(s)}^{\hat{c}},\lambda_{(s)}^{\hat{c}})$ & \tiny\Yvcentermath1$\yng(1,1)$ & $\bullet$ & $\bullet$ & $(1,3)'$\tabularnewline
\hline 
$(\mu_{(s)},h_{(s)}),(\mu'_{(s)},h'_{(s)})$ & \tiny\Yvcentermath1$\yng(1)$, \tiny\Yvcentermath1$\overline{\yng(1)}$ & $\bullet$ & \tiny\Yvcentermath1$\overline{\yng(1)}$ , \tiny\Yvcentermath1$\yng(1)$ & $(1,1),(1,1)$\tabularnewline
\hline 
$(a_{(g)}^{\mu},M_{(g)}^{\mu})$ & $\bullet$ & \tiny\Yvcentermath1$\yng(2)$ & $\bullet$ & $(2,2)$\tabularnewline
\hline 
$(D_{(g)}^{\hat{c}},\lambda_{(g)}^{\hat{c}})$ & $\bullet$ & \tiny\Yvcentermath1$\yng(1,1)$ & $\bullet$ & $(1,3)'$\tabularnewline
\hline 
$(\mu_{(g)},h_{(g)}),(\mu'_{(g)},h'_{(g)})$ & $\bullet$ & \tiny\Yvcentermath1$\yng(1)$ , \tiny\Yvcentermath1$\overline{\yng(1)}$ & \tiny\Yvcentermath1$\overline{\yng(1)}$ , \tiny\Yvcentermath1$\yng(1)$ & $(1,1),(1,1)$\tabularnewline
\hline 
$(w_{(g)}^{\dot{\alpha}},\mu_{(g)}^{\dot{\alpha}}),(w'{}_{(g)}^{\dot{\alpha}},\mu'{}_{(g)}^{\dot{\alpha}})$ & $\bullet$ & \tiny\Yvcentermath1$\yng(1,1)$ & \tiny\Yvcentermath1$\yng(1,1)$ & $(\frac{1}{2},1),(\frac{1}{2},1)$\tabularnewline
\hline 
$(C_{(g)}^{\alpha},M_{(g)}^{\alpha}),(C'{}_{(g)}^{\alpha},M'{}_{(g)}^{\alpha})$ & $\bullet$ & \tiny\Yvcentermath1$\yng(2)$ & $\bullet$ & $(1,1),(1,1)$\tabularnewline
\hline 
$(C_{(gs)}^{\alpha},M_{(gs)}^{\alpha}),(C'{}_{(gs)}^{\alpha},M'{}_{(gs)}^{\alpha})$ & \tiny\Yvcentermath1$\yng(1)$ , \tiny\Yvcentermath1$\overline{\yng(1)}$ & \tiny\Yvcentermath1$\overline{\yng(1)}$ , \tiny\Yvcentermath1$\yng(1)$  & $\bullet$ & $(1,1),(1,1)$\tabularnewline
\hline 
$(\chi_{(g)}^{\dot{\alpha}},\lambda_{(g)}^{\dot{\alpha}}),(\chi'{}_{(g)}^{\dot{\alpha}},\lambda'{}_{(g)}^{\dot{\alpha}})$ & $\bullet$ & \tiny\Yvcentermath1$\yng(1,1)$  & $\bullet$ & $(\frac{1}{2},1),(\frac{1}{2},1)$\tabularnewline
\hline 
$(\chi_{(gs)}^{\dot{\alpha}},\lambda_{(gs)}^{\dot{\alpha}}),(\chi'{}_{(gs)}^{\dot{\alpha}},\lambda'{}_{(gs)}^{\dot{\alpha}})$ & \tiny\Yvcentermath1$\yng(1)$ , \tiny\Yvcentermath1$\overline{\yng(1)}$ & \tiny\Yvcentermath1$\overline{\yng(1)}$ , \tiny\Yvcentermath1$\yng(1)$  & $\bullet$ & $(\frac{1}{2},1),(\frac{1}{2},1)$\tabularnewline
\hline 
\end{tabular}
\caption{All symmetry groups in the theory and the representations of the BRST pairs is demonstrated.  }
\end{center}
\end{table}

The integrals above get most simplified if one uses the basis of the Cartan subalgebras of the symmetry groups
in which the determinant of the $Q_\Omega$ squared should be calculated.
Using equations (\ref{QgQs})-(\ref{Wg}) and table 3 one can write the integral for the gauge instanton partition function as:
\begin{equation}\label{Zg}
 Z_{k}^{(g)}=\int\,\,\prod_{I=1}^{r[U\left(k\right)]}(\frac{d\tilde{\chi}_{I}}{2\pi i}) \Delta(\tilde{\chi}_{I})
 \frac{\mathcal{P}_{(g)}(\tilde{\chi_I})\mathcal{R}_{(g)}(\tilde{\chi_I})
 \mathcal{C}_{(g)}(\tilde{\chi_I})}{\mathcal{Q}_{(g)}(\tilde{\chi_I})\mathcal{L}_{(g)}(\tilde{\chi_I})\mathcal{W}_{(g)}(\tilde{\chi_I})}
\end{equation}
where $r[U\left(k\right)]$ is the rank of the gauge instanton group, $\Delta(\tilde{\chi}_{I})$ is the Vandermonde determinant which 
is in fact the Jacobi factor for going from the former basis to the Cartan basis. The functions in the integrand are expressed in
terms of the weight and root vectors of the associate symmetry groups.
Let us introduce all the integrand functions in the new basis. The Vandermonde (Jacobi) determinant is given by 
\begin{equation}\label{Deltag}
 \Delta(\tilde{\chi}_{I})=\prod_{\overrightarrow{\tilde\rho}\in adj\neq0}^{r[U(k)]}\overrightarrow{\tilde{\chi}}.
 \overrightarrow{\rho}=\prod_{I<J}^{k}(\tilde{\chi}_{I}-\tilde{\chi}_{J})^2
\end{equation}
where $\overrightarrow \rho$ is the root vector of the group U$(k)$ in the adjoint representation. See appendix \ref{WR} for 
weights and roots of the group U$(k)$ in different representations. The function $\mathcal P(\tilde\chi_I)$
is given by the following expression
\begin{equation}
 \mathcal{P}_{(g)}\left(\tilde{\chi_I}\right)=\prod_{\overrightarrow{\tilde{\rho}}\in adj}
 \left(\overrightarrow{\tilde{\chi}}.\overrightarrow{\tilde{\rho}}-\epsilon\right)=\prod_{I<J}^{k}
 \left(-\epsilon\right)^{k+1}\left[\left(\tilde{\chi}_{I}-\tilde{\chi}_{J}\right)^{2}-\epsilon^2\right]
\end{equation}
where $\epsilon=E_1+E_2$ with $E_A, A=1,2$ are defined below. The function $\mathcal {R}_{(g)}(\tilde\chi)$ involves the \emph{vev} of scalar field $\phi$, the bosonic
part of the chiral superfield. It is given by
\begin{equation}
 \mathcal{R}_{(g)}\left(\tilde{\chi_I}\right)=\prod_{\overrightarrow{\pi}\in F}\prod_{\overrightarrow{\gamma}\in F}
 \left(\overrightarrow{\tilde{\chi}}.\overrightarrow{\pi}+\overrightarrow{\phi}.\overrightarrow{\gamma}\right)=
 \prod_{I=1}^{k}\prod_{l=1}^{N}\left(\tilde{\chi}_{I}+\phi_{l}\right)
 \end{equation}
with $\overrightarrow \pi$ being the weight vector in the fundamental representation of the gauge 
instanton group U$(k)$ and $\overrightarrow \gamma$ 
the weight in the fundamental representation of the gauge group U$(N)$. 
The function $\mathcal Q_{(g)}(\tilde\chi_I)$ associates with the determinant of BRST charge squared of the moduli $a_{(g)}^\mu$. As
$a^\mu_{(g)}$ is in the antisymmetric representation of the gauge instanton group U$(k)$, in the Cartan basis it takes the following form:
\begin{equation}
\mathcal{Q}_{(g)}(\tilde\chi_I) =\prod_{\overrightarrow{\sigma}\in asym}
 \prod_{\overrightarrow{\beta}\in V}^{\left(+\right)}(\overrightarrow{\tilde{\chi}}.\overrightarrow{\tilde{\sigma}}
 -\overrightarrow{f}.\overrightarrow{\beta})=(E_{1}E_{2})^{k}\prod_{I<J}^{k}\prod_{A=1}^{2}[(\tilde{\chi}_{I}
 -\tilde{\chi}_{J})^{2}-E_{A}^{2}]
\end{equation}
where $\overrightarrow\sigma$ stands for the weights of the U$(k)$ group in the antisymmetric representation,
$\overrightarrow\beta$ is the positive weight of the twisted group SU$(2)\times$SU$(2)'$ and $E_A$ with $A=1,2$ 
are the eigenvalues of graviphoton field strength matrix. 
The determinant of $Q_\Omega^2$ acting on the charged bosonic modulus $w^{\dot\alpha}$ involves the \emph{vev} of the scalar field 
as well as the field strength of the graviphoton background. As the pairs $(w^{\dot\alpha},\mu^{\dot\alpha})$ and 
$(\bar w_{\dot\alpha},\bar\mu_{\dot\alpha})$ belong to the fundamental representations of the groups U$(k)$ and U$(N)$
the integral (\ref{Wg}) reads:
\begin{equation}
\mathcal{W}_{(g)}(\tilde\chi_I)=\prod_{\overrightarrow{\pi}\in F}\prod_{\overrightarrow{\gamma}\in F}
\left(\overrightarrow{\tilde{\chi}}.\overrightarrow{\pi}-\overrightarrow{\phi}.\overrightarrow{\gamma}
-\overrightarrow{f}.\overrightarrow{\beta}\right)=\prod_{I}^{k}\prod_{l}^{N}\left(\tilde{\chi}_{I}-\phi_{l}+\epsilon\right)
  \end{equation}
The functions $\mathcal{W}_{(g)}(\tilde\chi_I)$ and $\mathcal{R}_{(g)}\left(\tilde{\chi}\right)$ appear 
in the integrand of the gauge instanton partition function due to respectively bosonic and 
fermionic moduli states stretching between gauge branes and D-instantons on node 2 (3). 
Note that the orientifold projection acts similarly on the gauge branes and D-instantons sitting on node 2, because the matrix representation of the 
orientifold group on node 2 is antisymmetric and is given by (\ref{gamma-}). However the matrix representation of the orientifold acting on states on node 1
is symmetric (see equation (\ref{gamma})). 
The function $\mathcal{L}_{(g)}\left(\tilde{\chi_I}\right)$ and $\mathcal{C}_{(g)}\left(\tilde{\chi_I}\right)$ are linked to 
bosonic and fermionic string states between D-instantons on node 2 only. 
They are therefore in the bi-fundamental representation of the group U$(k)$:
\begin{equation}
 \mathcal{C}_{(g)}\left(\tilde{\chi_I}\right)=\prod_{\overrightarrow{\zeta}\in sym}
 \prod_{\overrightarrow{\beta}\in V}(\overrightarrow{\tilde{\chi}}.\overrightarrow{\zeta}+\overrightarrow{f}.\overrightarrow{\beta})
 =\prod_{I<J}^{k}\prod_{A}^{2}\left(2\tilde{\chi}_{I}+E_{A}\right)\left(\tilde{\chi}_{I}+\tilde{\chi}_{J}+E_{A}\right)
\end{equation}
\begin{equation}
 \mathcal{L}_{(g)}\left(\tilde{\chi_I}\right)=\prod_{\overrightarrow{\zeta}\in asym}(\overrightarrow{\tilde{\chi}}.\overrightarrow{\zeta}+f)
 =\prod_{I<J}^{k}[(\chi_I+\chi_J)^2-f^2]
\end{equation}
Putting all functions together in equation (\ref{Zg}) the partition function for gauge $k$-instanton in $\mathcal N=2$ U$(N)$ with a 
matter multiplet in symmetric representation reads
\begin{equation}
\begin{split}
 Z^{(g)}_k=
 \frac{\epsilon^k}{(E_1E_2)^k}\int\prod_{I=1}^{k}d\tilde\chi_I\prod_{l=1}^{N}\prod_{A=1}^{2}
 \frac{(2\tilde\chi_I+E_A)(\tilde\chi_I+\phi_l)}{(\tilde\chi_I+\phi_l-\epsilon)(\tilde\chi_I+\phi_l+\epsilon)}\\
\times  \frac{(\tilde\chi_I-\tilde\chi_J)^2[(\tilde\chi_I-\tilde\chi_J)^2-\epsilon^2](\tilde\chi_I+\tilde\chi_J+E_A)}
  {[(\tilde\chi_I-\tilde\chi_J)^2-E_A^2][(\tilde\chi_I+\tilde\chi_J)^2-\epsilon^2]}
 \end{split}
\end{equation}
This result has been obtained already by ADHM instanton construction (see e.g. \cite{Shadchin:2005mx} and \cite{Marino:2004cn} for the results
with different conventions).  

The partition function for stringy instanton with charge $k$ is given by:
\begin{equation}
  Z_{k}^{(s)}=\int \prod_{i=1}^{r[SO\left(k\right)]}(\frac{d{\chi}_{i}}{2\pi i}) \Delta({\chi}_{i})
 \frac{\mathcal{P}_{(s)}({\chi_i})\mathcal{R}_{(s)}({\chi_i})
}{\mathcal{Q}_{(s)}({\chi_i})}
\end{equation}
where $r[SO\left(k\right)]$ is the rank of group SO$(k)$. 
All functions $\mathcal{P}_{(s)}({\chi_i}), \mathcal{R}_{(s)}({\chi_i})$ and $\mathcal{Q}_{(s)}({\chi_i})$ 
are defined similarly as in the gauge instanton partition function.
The difference is that now the weight vectors should be calculated in the representations of the stringy instanton group, i.e. SO$(k)$. For functions  
$\mathcal{P}_{(s)}({\chi_i})$ and $\mathcal{R}_{(s)}({\chi_i})$ the weights are in the same representations as their gauge partner
i.e. respectively in the adjoint representation of SO$(k)$ and bifundamental representation of the groups 
SO$(k)$ and U$(N)$. However the weight vector in the function $\mathcal{Q}_{(s)}({\chi_i})$ as opposed to its gauge partner 
should be in the symmetric representation of the group SO$(k)$. These functions have been elaborated in appendix A of \cite{Ghorbani:2011xh}. 
The stringy partition function for odd and even instanton charge $k$ is given by:\\
$k\equiv odd$
\begin{equation}
\begin{split}
 Z_k^{(s)}=\frac{(-\epsilon)^{(k-1)/2}}{\epsilon^{(k+1)/2}}\int \prod_{i=1}^{(k-1)/2}d\chi_i
 \prod_{A=1}^2\prod_{l=1}^{N}\frac{\phi_l(\chi_i^2-\phi_l^2)^2 (\chi_i^2-\epsilon^2)} 
 {(\chi_i^2-E_A^2)(4\chi_i^2-E_A^2)} \\
 \times \prod_{i<j=1}^{(k-1)/2}\frac{[(\chi_i^2-\chi_j^2)^2[(\chi_i+\chi_j)^2-f^2][(\chi_i-\chi_j)^2-\epsilon^2] }
 {[(\chi_i+\chi_j)^2-E_A^2][(\chi_i-\chi_j)^2-E_A^2]}
  \end{split}
\end{equation}
$k\equiv even$
\begin{equation}
\begin{split}
 Z_k^{(s)}=\frac{(-\epsilon)^{k/2}}{\epsilon^{k/2}}\int \prod_{i=1}^{k/2} d\chi_i \prod_{A=1}^2 \prod_{l=1}^{N}
 \frac{(\chi_i^2-\phi_l^2)^2}{(4\chi_i^2-E_A^2)}\\
 \times \prod_{i<j=1}^{k/2}\frac{(\chi_i^2-\chi_j^2)^2[(\chi_i+\chi_j)^2-\epsilon^2][(\chi_i-\chi_j)^2-\epsilon^2]}
 {[(\chi_i+\chi_j)^2-E_A^2][(\chi_i-\chi_j)^2-E_A^2]}
 \end{split}
\end{equation}

The gauge-stringy instanton partition function in addition to the gauge and stringy parts consists 
also of a gauge-stringy part which includes the moduli $\mathcal C(\chi_i,\chi_I)$ and $\mathcal L(\chi_i,\chi_I)$. We have
\begin{equation}
 Z^{(gs)}_k=\int \prod_{i=1}^{r[SO\left(k\right)]}\prod_{I=1}^{r[U\left(k\right)]}
 (\frac{d{\chi}_{i}}{2\pi i}) (\frac{d\tilde{\chi}_{I}}{2\pi i})
 \Delta({\chi}_{i}) \Delta(\tilde{\chi}_{I})
  \mathcal{I}_{(s)}(\chi_i) \mathcal{I}_{(g)}(\tilde\chi_I) \mathcal{I}_{(gs)}(\chi_i,\tilde\chi_I)
 \end{equation}
 where 
 \begin{equation}
  \mathcal{I}{(\chi_i,\tilde\chi_I)}=\frac{\mathcal C(\chi_i,\chi_I)}{\mathcal L(\chi_i,\chi_I)}.
 \end{equation}
$\Delta({\chi}_{i})$ and $\Delta(\tilde{\chi}_{I})$ are defined respectively in (\ref{Deltag}) and appendix A in \cite{Ghorbani:2010ks}. Also: \\
$k\equiv odd$
\begin{equation}
\mathcal C(\chi_i,\chi_I)=-\prod_{i=1}^{(k-1)/2}\prod_{I=1}^{k}\prod_{A=1}^{2}\left[\left(\chi_{i}
 +\tilde{\chi}_{I}\right)^{2}-E_{A}^{2}\right]\left[\left(\chi_{i}
 -\tilde{\chi}_{I}\right)^{2}-E_{A}^{2}\right]\left(\tilde{\chi}_{I}^{2}-E_{A}^{2}\right)
\end{equation}
\begin{equation}
\mathcal L(\chi_i,\chi_I)=-\prod_{i=1}^{(k-1)/2}\prod_{I=1}^{k}
\left[\left(\chi_{i}+\tilde{\chi}_{I}\right)^{2}-f^{2}\right]\left[\left(\chi_{i}
-\tilde{\chi}_{I}\right)^{2}-f^{2}\right]\left[\tilde{\chi}_{I}^{2}-f^{2}\right]
\end{equation}
$k \equiv even$
\begin{equation}
 \mathcal C(\chi_i,\chi_I)=\prod_{i=1}^{k/2}\prod_{I=1}^{k}\prod_{A=1}^{2}
\left[\left(\chi_{i}+\tilde{\chi}_{I}\right)^{2}-E_{A}^{2}\right]
\left[\left(\chi_{i}-\tilde{\chi}_{I}\right)^{2}-E_{A}^{2}\right]
\end{equation}
\begin{equation}
  \mathcal L(\chi_i,\chi_I)=\prod_{i=1}^{k/2}\prod_{I=1}^{k}
  \left[\left(\chi_{i}+\tilde{\chi}_{I}\right)^{2}-f^{2}\right]
  \left[\left(\chi_{i}-\tilde{\chi}_{I}\right)^{2}-f^{2}\right]
\end{equation}
The gauge-stringy $k$-instanton partition function then takes the following form:\\
$k \equiv odd$
\begin{equation}
\begin{split}
 Z^{(gs)}_k=\int \prod_{i=1}^{(k-1)/2}\prod_{I=1}^{k}
 (\frac{d{\chi}_{i}}{2\pi i}) (\frac{d\tilde{\chi}_{I}}{2\pi i})
 \frac{\chi_i^2 (\chi_i^2-\chi_j^2)^2 (\tilde\chi_I-\tilde\chi_J)^2(\tilde{\chi}_{I}^{2}-E_{A}^{2})}{(\tilde{\chi}_{I}^{2}-\epsilon^{2})}\\
 \times \prod_{i=1}^{(k-1)/2} \prod_{=1}^{k} \frac{\left[\left(\chi_{i}
 +\tilde{\chi}_{I}\right)^{2}-E_{A}^{2}\right]\left[\left(\chi_{i}
 -\tilde{\chi}_{I}\right)^{2}-E_{A}^{2}\right]}
 {\left[\left(\chi_{i}+\tilde{\chi}_{I}\right)^{2}-\epsilon^{2}\right]\left[\left(\chi_{i}
-\tilde{\chi}_{I}\right)^{2}-\epsilon^{2}\right]}
\end{split}
\end{equation}
$k \equiv even$
\begin{equation}
\begin{split}
 Z^{(gs)}_k=\int \prod_{i=1}^{(k-1)/2}\prod_{I=1}^{k}
 (\frac{d{\chi}_{i}}{2\pi i}) (\frac{d\tilde{\chi}_{I}}{2\pi i})
(\chi_i^2-\chi_j^2)^2 (\tilde\chi_I-\tilde\chi_J)^2\\
 \times \prod_{i=1}^{(k-1)/2} \prod_{I=1}^{k} \frac{\left[\left(\chi_{i}
 +\tilde{\chi}_{I}\right)^{2}-E_{A}^{2}\right]\left[\left(\chi_{i}
 -\tilde{\chi}_{I}\right)^{2}-E_{A}^{2}\right]}
 {\left[\left(\chi_{i}+\tilde{\chi}_{I}\right)^{2}-\epsilon^{2}\right]\left[\left(\chi_{i}
-\tilde{\chi}_{I}\right)^{2}-\epsilon^{2}\right]}
\end{split}
\end{equation}


\section{Prepotential Corrections}\label{sec-pre}
The integration of prepotential over the superspace coordinates gives rise to $\mathcal N=2$ effective action. 
The gauge-stringy prepotential corrections in the presence of the $\Omega$-background that is the Ramond-Ramond 3-form in our
model is given by the logarithm of the total partition function:
\begin{equation}\label{FZ}
 F^{(n.p.)}(\Phi)=\epsilon \log \mathcal Z_{tot}\mid_{\phi\to\Phi,E_A\to 0} 
\end{equation}
where $\mathcal Z_{tot}$ is given by
\begin{equation}
 \mathcal{Z}_{tot}=\sum_k Z_k q^k.
\end{equation}
The prepotential can be expanded in powers of $q$:
\begin{equation}\label{FFk}
 F^{(n.p.)}=\sum_{k=1}^{\infty}F_k q^k \mid_{\phi\to\Phi,E_A\to 0} 
\end{equation}
where $q=\exp(2\pi i \tau)$ is dimensionless. 
Uning (\ref{FZ}) and (\ref{FFk}) the prepotential factors $F_k$ are related to the partition functions $Z_k$. For 
$k=1,2$ we have,
\begin{equation}\label{F1Z1F2Z2}
\begin{split}
 & F_1^{(gs)}=\epsilon Z_1^{(gs)}\\
 & F_2^{(gs)}=\epsilon Z_2^{(gs)} -{F_1^{(gs)}}^2/2\epsilon.
 \end{split}
\end{equation}
In the partition function integrations we obey the requirement that $E_1+E_2=0$ (see e.g. \cite{Shadchin:2005mx}) which should be held 
for the eigenvalues of the field strength  of the graviphoton background. 
The result of the $k=1$ partition function for an arbitrary gauge group U$(k)$ turns out to be
\begin{equation}\label{Z1gs}
 Z_1^{(gs)}=\frac{(-1)^N}{2} \mathcal N_1 E_1^2,
\end{equation}
and for $k=2$ we have,
\begin{equation}\label{Z2gs}
 Z_2^{(gs)}=\frac{(-1)^N}{3} \mathcal N_2 E_1^4\left( 8 \text{tr}\Phi^4-4 E_1^2 \text{tr}\Phi^2+5/16 E_1^4 \right)
\end{equation}
where $\mathcal N_1$ and $\mathcal N_2$ are overall factors.  As seen from equations (\ref{Z1gs}) and (\ref{Z2gs}) 
the first two partition functions do not have any singularity in the $E_1$ and $E_2$ after setting $E_1+E_2=0$. 
The prepotential corrections given by (\ref{F1Z1F2Z2}) are the following
\begin{equation}
 F_1^{(gs)}=\frac{(-1)^N}{2} \mathcal N_1 E_1^4
\end{equation}
and
\begin{equation}
 F_2^{(gs)}=\frac{(-1)^N}{3} \mathcal N_2 E_1^6\left( 8 \text{tr}\Phi^4-4 E_1^2 \text{tr}\Phi^2+5/16 E_1^4 \right)
 -\frac{\mathcal N_1^2}{8}  E_1^6
\end{equation}
Again clearly no singularity occurs in $F_1$ and $F_2$ to be fixed by coefficients $\mathcal N_1$ and $\mathcal{N}_2$. 
Moreover although in the presence of the graviphoton field strength  the corrections depend on the chiral superfield $\Phi$, 
in the flat limit where $E_1,E_2\to 0$, at least the first two gauge-stringy instanton corrections are vanishing.
The calculations to take the counter integrals for $k>2$ become very involved and we refrain from doing that.

\section{Conclusions}\label{sec-con}
In this work we have considered the well-studied D3/D(--1) brane set-up at the singularity of the orbifold 
$\mathbb C^2/\mathbb Z_3\times\mathbb C$ in the presence of an O3-plan. In this set-up we have introduced
the gauge-stringy instantons in a particular supersymmetry theory i.e. the $\mathcal N=2$ U$(N)$ gauge theories
with one matter multiplet in the symmetric representation of the gauge group U$(N)$. The gauge-stringy set-ups defined here are simply 
the combinations of the gauge and stringy instanton set-ups in D3/D(--1) brane system. We remind that the gauge instanton configurations in D3/D(--1) branes 
corresponds to the arrangement of D3 and D(--1) branes such that they lie on the same representation of the orbifold groups. In the quiver diagram Fig. 1
it means that both D3 and D(--1) stacks of branes are put e.g. on node 2. The stringy instanton arrangement is the situation that D3 and D(--1) branes 
are put in different representation of orbifold group which means having D(--1) branes e.g. on node 1 and D3 branes on node 2 in the quiver diagram. 
In this work we have assumed that the stacks of D(--1) branes can occupy both nodes 1 and 2 (and therefore node 3 due to the presence of the orientifold plane).
We however choose only one stack of gauge branes D3 to be on node 2 (3) because we are interested in having a simple gauge group U$(N)$. Clearly 
the model can be more complicated by considering gauge branes also on node 1, That would have led to gauge-stringy instantons in U$(N)\times$SO$(N')$ gauge group.
But what is important in gauge-stringy models, is the presence of D-instantons in different representations of the orbifold group.  
Our focus in the current work has been on gauge-stringy effects in the effective action with gauge group U$(N)$. Having chosen gauge-stringy
configuration it is easily seen that in addition to the gauge and stringy moduli there are extra states that can
be found neither in gauge nor in stringy moduli. These are in fact D(--1)/D(--1) bosonic and fermionic neutral states. We have shown that the new zero-modes 
do not change the length dimension of the moduli measure, although they do contribute in the corrections of the prepotential. The dimension of the moduli
measure turns out to be 
\begin{equation}
 [d\mathcal M_{(tot)}]=(N-2)(k_g-k_s)
\end{equation}
We see that the moduli measure happens to be dimensionless in two cases. When $N=2$ i.e. choosing gauge group SU$(2)$ 
regardless of the values of gauge and stringy charges $k_g$ and $k_s$ the measure is dimensionless. This case which was studied in details in \cite{Ghorbani:2010ks}
is a superconformal theory.
Another choice is when $k_g=k_s$ i.e. when the gauge and stringy instanton charges are equal. Having satisfied this condition in our model, for any gauge group
U$(N)$ the measure is dimensionless. In this paper we have set $k_g=k_s\equiv k$ and calculated the leading corrections to the prepotential due to $k=1,2$ 
gauge stringy instanton. It turns out that in the zero limit of the graviphoton background the gauge-stringy instanton corrections for $k=1,2$ are vanishing. 
The higher order corrections are very involved and it is not obvious if this result can be true for all instanton charges. 

We have built our brane set-up at the singularity of a specific orbifold group, $\mathbb Z_3$. The reason is that we needed to 
merge stringy instantons with unusual instanton group SO$(k)$ and ordinary instantons in one single model. Choosing 
a certain representation of the orientifold plan on one node of the quiver allows us to project U$(k)$ instantons 
into O$(k)$ instantons while leaving U$(k)$ instantons on the other nodes unchanged. In the end of the day one 
is left with a model consisting of both ordinary and stringy instantons and their interactions which are described by the new 
zero modes called gauge-stringy neutral moduli in this paper. 
It is however interesting to find sophisticated models at the singularity 
of the more general orbifold, $\mathbb C^2/\mathbb Z_{2m+1}$. In such a background the number 
of nodes in the quiver diagram increase to $2m+1$. Therefore to build the same model which combines the instantons with instanton groups O$(k)$ and U$(k)$ 
one should be careful about the group representations of the orientifold acting on different nodes. 
Filling some nodes of $\mathbb C^2/\mathbb Z_{2m+1}$ quiver with gauge branes and D-instantons together with appropriate orientifold plans may provide 
us with more general gauge-stringy models with larger gauge and instanton groups.  

It is also worthy to investigate if gauge-stringy configurations may lead to a dimensionless moduli measure also in other supersymmetry theories. In particular 
it is interesting to verify the vanishing non-perturbative corrections for other gauge-stringy models.

\section*{Acknowledgments}
I am specially grateful to Alberto Lerda and Niclas Wyllard for very fruitful comments and discussions. I would like to 
thank Daniele Musso for many useful conversations we had about instantons in string theory during my stay in Torino. 
I would also like to thank Marco Bill\'o and Michael Wohlgenannt for discussions.

 
 \appendix
\section{\texorpdfstring{$\left(\Sigma^{p}\right)^{AB}$ and $\left(\bar{\Sigma}^{p}\right)_{AB}$ in Terms of Pauli Matrices in 4d}{Lg}}\label{SigmaSigmabar}
The matrices $\left(\Sigma^{p}\right)^{AB}$ and $\left(\bar{\Sigma}^{p}\right)_{AB}$ are similar to the Pauli matrices 
in 4-dimensions being the components of gamma matrix in 6-dimensions:
\begin{equation}
\Gamma^p=\left(\begin{array}{cc}
0 & \Sigma^{p}\\
\bar{\Sigma}^{p} & 0
\end{array}\right). 
\end{equation}
The gamma matrix should satisfy the Clifford algebra hence the sigma matrices are expressible by the 
't Hooft (anti-)selfdual matrices. See for example appendix A in \cite{Billo:2002hm}. 
$\left(\Sigma^{p}\right)^{AB}$ and $\left(\bar{\Sigma}^{p}\right)_{AB}$ however can be presented 
in terms of the Pauli matrices in 4-dimensions which is more convenient for our purposes:
\begin{equation}\label{Sigma}
 \left(\Sigma^{p}\right)^{AB}=\left(\begin{array}{cc}
\epsilon^{ab}\left(\delta_{8}^{p}-i\delta_{9}^{p}\right) & \left(\bar{\sigma}^{m}\right)^{\dot a b}\delta_{m}^{p}\\
\left(\sigma^{m*}\right)^{a\dot b}\delta_{m}^{p} & \epsilon^{\dot{a}\dot{b}}\left(\delta_{8}^{p}+i\delta_{9}^{p}\right)
\end{array}\right)
\end{equation}

\begin{equation}\label{Sigmabar}
 \left(\bar{\Sigma}^{p}\right)_{AB}=\left(\begin{array}{cc}
\epsilon_{ab}\left(\delta_{8}^{p}+i\delta_{9}^{p}\right) & \left(\bar{\sigma}^{m*}\right)_{\dot a b}\delta_{m}^{p}\\
\left(\sigma^{m}\right)_{a \dot b}\delta_{m}^{p} & \epsilon_{\dot{a}\dot{b}}\left(\delta_{8}^{p}-i\delta_{9}^{p}\right)
\end{array}\right)
\end{equation}
where $m=4,5,6,7$ is the first four real coordinates in the internal space and,
\begin{equation}
 \left(\sigma^{m}\right)_{a\dot b}=\left(\tau^{c},i\one \right)_{a \dot b}
~~~~~~~~~
 \left(\bar{\sigma}^{m}\right)_{\dot a b}=\left(-\tau^{c},i\one\right)_{\dot a b}
\end{equation}
with $\tau^{c}$ being the Pauli matrices in 4-dimensions. The following equalities hold for bared and unbared sigma matrices:
\begin{equation}
 \left(\bar{\sigma}^{m*}\right)_{\dot a b}=-\left(\sigma^{m}\right)_{b\dot a}
 ~~~~~~~~~
  \left({\sigma}^{m*}\right)_{a \dot b}=-\left(\bar\sigma^{m}\right)_{\dot b a}.
\end{equation}

\section{ \texorpdfstring{$Q$-Exact Moduli Action}{Lg}}
In this appendix we briefly explain how the moduli action of the $\mathcal N=2$ U$(N)$ can be expressed in terms of a singlet supercharge 
$Q$-exact action. The singlet supercharge can be obtained out of the doublet supersymmetry charges by imposing a topological-twist between a 
Lorentz SU$(2)$ subgroup and the internal SU$(2)$. Then we switch on a graviphoton background and redefine the BRST charge transformations by studying
the moduli interactions with graviphoton field. Finally we show that in a certain localization limit one can get rid of all non-Gaussian terms in the integrand of 
the partition function.    
\subsection{Moduli Action}\label{action}
We take the action from \cite{Billo:2002hm} by slightly modifying the definition of the auxiliary modulus $D^c$. The total moduli action can be decomposed as:
\begin{equation}
 S=S_{\text{cubic}}+S_{\text{quartic}}+S_{\text{charged}}.
 \end{equation}
The action in quartic part is: 
\begin{equation}
  g_{0}^{2}~S_{\text{quartic}}=-\frac{1}{4}\left[{a}^{\mu},{a}^{\nu}\right]^{2}
  -\frac{1}{2}\left[{a}_{\mu},{\chi}^{p}\right]^{2}-\frac{1}{4}\left[{\chi}_{p},{\chi}_{q}\right]^{2}
\end{equation}
where $\mu,\nu$ are the Lorentz indices and $p,q=1,..,6$.
We define the auxiliary modulus $D^c$ as follows:
 \begin{equation}
 D^c=-\frac{1}{2}\bar{\eta}_{\mu\nu}^{c}\left[a^{\mu},a^{\nu}\right]-\frac{1}{2}\bar{\zeta}_{mn}^{c}\left[\chi^{m},\chi^{n}\right]
 \end{equation}
 with $\bar\eta_{\mu\nu}^c$ being the 't Hooft anti-selfdual matrix (see the appendix in \cite{'tHooft:1976fv}) 
 and $\bar{\zeta}_{mn}^{c}$ a two-tensor satisfying
 \begin{equation}
 \bar{\zeta}_{mn}^{c}~\bar\eta_{\mu\nu}^c=0.
\end{equation}
The quartic part of the action then takes the form:
\begin{equation}\label{quartic}
\begin{split}
  g_0^2~S_{\text{quartic}}= \frac{1}{2}{D}_{c}^{2}+\frac{1}{2}{D}_{c}\left(\bar{\eta}_{\mu\nu}^{c}\left[{a}^{\mu},{a}^{\nu}\right]
  +\bar{\zeta}_{mn}^{c}\left[\chi^{m},\chi^{n}\right]\right)
  -\frac{1}{4}\left[{a}_{\mu},{\chi}\right]\left[{a}_{\mu},{\bar{\chi}}\right]\\
  -\frac{1}{2}\left[a^\mu,\chi^{\dot\alpha a} \right]\left[a_\mu,\chi_{a \dot\alpha} \right] 
 -\frac{1}{4}\left[{\chi},{\bar{\chi}}\right]\left[{\chi},{\bar{\chi}}\right]
 -\frac{1}{4}\left[{\chi},\chi^{\dot\alpha a}\right]\left[{\bar{\chi}},\chi_{a \dot\alpha}\right],
  \end{split}
\end{equation}
where $n,m=4,..,7$ are the indices of the first four real coordinates in the internal space.

The cubic and charged parts of the action are the following:
\begin{equation}
\begin{split}
 g_{0}^{2}~S_{\text{cubic}}=i(\bar\sigma^\mu)_{\dot{\alpha}\beta} \left[{M}^{\beta A},{a}_{\mu}\right]\lambda_{\,\, A}^{\dot{\alpha}}
 -\frac{i}{2}\left(\Sigma^{p}\right)^{AB}{\lambda}_{\dot{\alpha}A}\left[{\chi}_{p},{\lambda}_{\,\, B}^{\dot{\alpha}}\right]\\
 -\frac{i}{2}\left(\bar{\Sigma}^{p}\right)_{AB}{M}^{\alpha A}\left[{\chi}_{p},{M}_{\alpha}^{\,\, B}\right] 
 \end{split}
\end{equation}
\begin{equation}\label{charged}
\begin{split}
 g_0^2~S_{\text{charged}}= 2i\left({\bar{\mu}}^{A}{w}_{\dot{\alpha}}
 +{\bar{w}}_{\dot{\alpha}}{\mu}^{A}\right)\lambda_{\,\, A}^{\dot{\alpha}}
 -2i{D}^{c}{\bar{w}}^{\dot{\alpha}}\left(\tau^{c}\right)_{\,\,\dot{\alpha}}^{\dot{\beta}}{w}_{\dot{\beta}}\\ 
 +2{\chi}{\bar{w}}_{\dot{\alpha}}{w}^{\dot{\alpha}}{\bar{\chi}}
  +2{\chi^m}{\bar{w}}_{\dot{\alpha}}{w}^{\dot{\alpha}}{\chi_m}
+i\left(\bar{\Sigma}^{p}\right)_{AB}{\bar{\mu}}^{A}{\mu}^{B}{\chi}_{p},
\end{split}
 \end{equation}
where the matrices $\left(\Sigma^{p}\right)^{AB}$ and $\left(\bar{\Sigma}^{p}\right)_{AB}$ has been introduced in appendix \ref{SigmaSigmabar}.
Using (\ref{Sigma}) and (\ref{Sigmabar}) the cubic and charged actions can be expressed as follows:
\begin{equation}\label{scubic}
\begin{split}
g_0^2~S_\text{cubic}=4(\bar\sigma^{\mu})_{\dot{\alpha}\beta}\left[{M}^{\beta a},{a}_{\mu}\right]\lambda_{\,\, a}^{\dot{\alpha}}
 +4(\bar\sigma^{\mu})_{\dot{\alpha}\beta}\left[{M}^{\beta\dot{a}},{a}_{\mu}\right]\lambda_{\,\,\dot{a}}^{\dot{\alpha}}\\
-\frac{i}{2}{\lambda}_{\dot{\alpha}a}\left[{\chi},{\lambda}^{\dot{\alpha}a}\right]
 -\frac{i}{2}{\lambda}_{\dot{\alpha}\dot{a}} \left[{\bar{\chi}},{\lambda}^{\dot{\alpha}\dot{a}}\right]
 -i{\lambda}_{\dot{\alpha} \dot{a}}\left[{\chi}^{\dot{a} {b}},{\lambda}_{\;\;{b}}^{\dot{\alpha}}\right]\\
 -\frac{i}{2}{M}^{\alpha a}\left[{\chi},{M}_{\alpha a}\right] -\frac{i}{2}{M}^{\alpha\dot{a}}\left[{\bar{\chi}},{M}_{\alpha \dot{a}}\right]
 -i {M}^{\alpha a}\left[{\chi}_{a \dot b},{M}_{\alpha}^{\,\, \dot b}\right]
   \end{split}
 \end{equation}
\begin{equation}\label{scharged}
\begin{split}
g_0^2~S_\text{charged}=2i\left({\bar{\mu}}^{a}{w}_{\dot{\alpha}}+{\bar{w}}_{\dot{\alpha}}{\mu}^{a}\right){\lambda}_{\,\, a}^{\dot{\alpha}}
 +2i\left({\bar{\mu}}^{\dot{a}}{w}_{\dot{\alpha}}
 +{\bar{w}}_{\dot{\alpha}}{\mu}^{\dot{a}}\right){\lambda}_{\,\,\dot{a}}^{\dot{\alpha}}\\
 -i{D}^{c}{\bar{w}}^{\dot{\alpha}}\left(\tau^{c}\right)_{\,\,\dot{\alpha}}^{\dot{\beta}}{w}_{\dot{\beta}}
 -{\chi}^{\dot a b}{\bar{w}}_{\dot{\alpha}}{w}^{\dot{\alpha}}{\chi}_{b \dot a}
 +2{\chi}{\bar{w}}_{\dot{\alpha}}{w}^{\dot{\alpha}}{\bar{\chi}}\\
 +i{\bar{\mu}}^{a}{\mu}_{a}{\chi}
 +i{\bar{\mu}}^{\dot{a}}{\mu}_{\dot{a}}{\bar{\chi}}
 +i\left({\bar{\mu}}^{a}{\mu}^{\dot b}-{\bar{\mu}}^{\dot b}{\mu}^{a}\right){\chi}_{a\dot b}
 \end{split}
 \end{equation}
 where we have defined,
 \begin{equation}
  \chi^{\dot a b}\equiv \left( \bar\sigma^m \right)^{\dot a b}\chi_m
  ~~~~~~~~~~~~~~~~~~\chi_{a \dot b}\equiv \left(\sigma^m \right)_{a \dot b}\chi_m.
 \end{equation}

In order to exploit the localization technique in multi-instanton calculus we need to construct the BRST structure of the action. 
First step toward this goal is to extract a BRST-like charge among four $\mathcal N=2$ supersymmetry charges. That is possible by making a topological twist
between e.g. the SU$(2)_R$ subgroup of the Lorentz group and a SU$(2)_I$ subgroup in the internal space:
\begin{equation}
 SU\left(2\right)_{R}\times SU\left(2\right)_{I}\rightarrow SU\left(2\right)'.
\end{equation}
This is equivalent to the substitution
\begin{equation}
 \dot a\rightarrow \dot{\alpha}
\end{equation}
in the expressions (\ref{scubic}) and (\ref{scharged}). As a result of the above substitution the moduli ${\lambda}_{\dot{\alpha} \dot a}$
and ${M}^{\alpha \dot a}$ decompose into
\begin{equation}
 {\lambda}_{\dot{\alpha} \dot a}\rightarrow {\lambda}_{\dot{\alpha}\dot{\beta}}=
\frac{i}{2}\left(\tau^{c}\right)_{\dot{\alpha}\dot{\beta}}{\lambda}_{c}+\frac{1}{2}\epsilon_{\dot{\alpha}\dot{\beta}}{\eta}
~~~~~~~~~~~
{M}^{\alpha \dot a}\rightarrow {M}^{\alpha\dot{\beta}}=\frac{1}{2}\left(\sigma^{\mu}\right)^{\alpha\dot{\beta}}{M}_{\mu}.
\end{equation}

After the topological twist the cubic and charged actions take the following form:
\begin{equation}
\begin{split}
 g_{0}^{2}S_{\text{cubic}}=i(\bar\sigma^{\mu})_{\dot{\alpha}\beta}\left[{M}^{\beta a},{a}_{\mu}\right]\lambda_{~a}^{\dot{\alpha}}
 +\frac{i}{4}\left[M^\mu,a_\mu \right]\eta
 +\frac{i}{4}{\lambda}_{c}\left[{\chi},{\lambda}^{c}\right]
 -\frac{i}{4}{\eta}\left[{\chi},{\eta}\right]\\
 -\frac{i}{2}{\lambda}_{\dot{\alpha} a}\left[{\bar{\chi}},{\lambda}^{\dot{\alpha} a}\right]
 +\frac{1}{2}\left(\tau^c\right)_{\dot\alpha \dot\beta}{\lambda}_c\left[{\chi}^{\dot\beta b},{\lambda}_{\;\;b}^{\dot{\alpha}}\right]
  +\frac{i}{2}\eta \left[{\chi}^{\dot\alpha b},{\lambda}_{\dot{\alpha}b}\right]\\
 -\frac{i}{4}{M}_{\mu}\left[{\bar{\chi}},{M}^{\mu}\right]
 -\frac{i}{2}{M}^{\alpha a}\left[{\chi},{M}_{\alpha a}\right]
 + 4\left(\bar\sigma^\mu\right)_{\alpha}^{~\dot \beta}{M}^{\alpha a}\left[{\chi}_{a \dot \beta},{M}_\mu\right]\\
 \end{split}
 \end{equation}
 \begin{equation}
 \begin{split}
 g_{0}^{2}S_{\text{charged}}=-\left({\bar{\mu}}^{\dot{\alpha}}{w}^{\dot{\beta}}
 +{\bar{w}}^{\dot{\alpha}}{\mu}^{\dot{\beta}}\right)\left(\tau^{c}\right)_{\dot{\alpha}\dot{\beta}}{\lambda}_{c}
 -\left({\bar{\mu}}^{\dot{\alpha}}{w}_{\dot{\alpha}}
 +{\bar{w}}_{\dot{\alpha}}{\mu}^{\dot{\alpha}}\right){\eta}\\
 -\left({\bar{\mu}}^{a}{w}^{\dot{\alpha}}+{\bar{w}}^{\dot{\alpha}}{\mu}^{a}\right){\lambda}_{a\dot{\alpha}}
 -{D}^{c}{\bar{w}}^{\dot{\alpha}}\left(\tau^{c}\right)_{\dot{\alpha}\dot{\beta}}{w}^{\dot{\beta}}
 +i{\bar{w}}^{\dot{\alpha}}{w}_{\dot{\alpha}}\left({\chi}{\bar{\chi}}+{\chi}{\bar{\chi}}\right)\\
 -{\chi}_{\dot{\alpha}a}{\bar{w}}^{\dot{\alpha}}{h}^{a}
 -{\bar{h}}^{a}{w}^{\dot{\alpha}}{\chi}_{a\dot{\alpha}}
 +{\bar{h}}^{a}{h}_{a}\\
 +2i{\bar{\mu}}^{\dot{\alpha}}{\mu}_{\dot{\alpha}}{\bar{\chi}}
 -i{\bar{\mu}}^{a}{\mu}_{a}{\chi}
 +\left({\bar{\mu}}^{a}{\mu}^{\dot \alpha}-{\bar{\mu}}^{\dot \alpha}{\mu}^{a}\right){\chi}_{a\dot\alpha}
 \end{split}
 \end{equation}

where we have made use of 
\begin{equation}
 \left(\bar\sigma_\mu\right)_{\dot\alpha \beta} \left(\sigma_\nu\right)^{\beta \dot\gamma}=-2\delta_{\mu\nu}\epsilon_{\dot\alpha}^{~\dot\gamma}
 ~~~~~~~~~~~~~
 \sigma_\mu^\dagger=-\bar\sigma_\mu
\end{equation}


\subsection{BRST Structure}
In this subsection we discuss on the BRST structure of the action with and without the graviphoton background. The graviphoton background 
(the massless RR string state) which interacts with some of the moduli play the r\^ole of a regulator in partition function integral. The graviphoton background
is in fact equivalent
to the $\Omega$ background introduced by Nekrasov in the localization technique. The action after the topological twist can be rewritten 
as a $Q$-exact expression, i.e. it can be written as the action of $Q$ on an expression:
\begin{equation}
 S=Q\bar{\Xi}
\end{equation}
where $Q=\epsilon^{\dot\alpha\dot\beta}Q_{\dot\alpha\dot\beta}$ and $\bar{\Xi}$ is the so-called \emph{gauge fermion} and is given by:
\begin{equation}
\begin{split}
 {\Xi}=\frac{i}{4}{M}^{\mu}\left[{\bar{\chi}},{a}_{\mu}\right]
 +\frac{1}{2}A\bar{\eta}_{\mu\nu}^{c}{\lambda}^{c}\left[{a}_{\mu},{a}_{\nu}\right]
 -{\bar{w}}^{\dot{\alpha}}\left(\tau^{c}\right)_{\dot{\alpha}\dot{\beta}}{w}^{\dot{\beta}}{\lambda}^{c}
 +\left({\bar{\mu}}^{\dot{\alpha}}{w}_{\dot{\alpha}}
 +{\bar{w}}_{\dot{\alpha}}{\mu}^{\dot{\alpha}}\right){\bar{\chi}}\\
 +\frac{1}{2}{\lambda}^{c}{D}^{c}
 +\frac{i}{4}\left[{\chi},{\bar{\chi}}\right]{\eta}
 -\frac{1}{2}\left({\bar{\mu}}^{a}{h}_{a}
 +{\bar{h}}^{a}{\mu}_{a}\right)
 -\left({\bar{w}}^{\dot{\alpha}}{\mu}^{a}+{\bar{\mu}}^{a}{w}^{\dot{\alpha}}\right){\chi}_{a\dot{\alpha}}\\
 +4\left(\bar{\sigma}^{\mu}\right)_{\,\,\alpha}^{\dot{\alpha}}\left[{\chi}_{a\dot{\alpha}},{a}_{\mu}\right]{M}^{\alpha a}
 +\frac{1}{2}{M}^{\alpha a}{C}_{\alpha a}
 -\frac{i}{2}{\lambda}_{\dot{\alpha}a}\left[{\chi}^{\dot{\alpha}a},{\bar{\chi}}\right]\\
 +\frac{1}{2}\bar{\zeta}_{mn}^{c}\left( \bar\sigma^m \sigma^n \right)^{~\dot\beta}_{\dot\alpha}{\lambda}_{c}\left[{\chi}^{a\dot{\alpha}},{\chi}_{\dot{\beta}a}\right]
 \end{split}
\end{equation}
The action of BRST charge $Q$ on various moduli is as follows,

\begin{equation}
 \begin{aligned}
 & Q{\chi}=0,~~Q{\bar{\chi}}={\eta}~~~~~\\
 & Q{a}^{\mu}={M}^{\mu}~~~~~\\
 & Q{\lambda}^{c}={D}^{c}~~~~~\\
 & Q{w}^{\dot{\alpha}}={\mu}^{\dot{\alpha}}~~~~~\\
 & Q{\bar{w}}_{\dot{\alpha}}={\bar{\mu}}_{\dot{\alpha}}~~~~~\\
 & Q{\chi}^{\dot\alpha a}={\lambda}^{\dot{\alpha} a}~~~~~\\
 & Q{\chi}_{a\dot\alpha}={\lambda}_{a\dot\alpha}~~~~~\\
 & Q{M}^{\alpha a}={C}^{\alpha a}~~~~~\\
 & Q{\mu}^{a}={h}^{a}~~~~~\\
 & Q{\bar{\mu}}^{a}={\bar{h}}^{a}~~~~~\\
 \end{aligned}
\begin{aligned}
& Q{\eta}=i\left[{\chi},{\bar{\chi}}\right]\\
& Q{M}^{\mu}=i\left[{\chi},{a}^{\mu}\right] \\
& Q{D}^{c}=i\left[{\chi},{\lambda}^{c}\right]\\
& Q{\mu}^{\dot{\alpha}}=-i{w}^{\dot{\alpha}}{\chi}\\
& Q{\bar{\mu}}_{\dot{\alpha}}=i{\chi}{\bar{w}}_{\dot{\alpha}}\\
& Q{\lambda}^{\dot{\alpha} a}=i\left[{\chi},{\chi}^{\dot{\alpha} a}\right]\\
& Q{\lambda}_{a \dot{\alpha}}=i\left[{\chi},{\chi}_{a \dot{\alpha}}\right]\\
& Q{C}^{\alpha a}=i\left[{\chi},{M}^{\alpha a}\right]\\
& Q{h}^{a}=i{\mu}^{a}{\chi}\\
& Q{\bar{h}}^{a}=-i{\chi}{\bar{\mu}}^{a}\\
\end{aligned}
\end{equation}


\subsection{BRST Structure in Graviphoton Background}
We switch on the Ramond-Ramond 3-form graviphoton field strength  $\mathcal F_{\mu\nu\rho}$ with holomorphic and anti-holomorphic 
components $f_{\mu\nu}\equiv \mathcal F_{\mu\nu z}$ and $\bar{f}_{\mu\nu}\equiv \mathcal F_{\mu\nu\bar{z}}$ 
($z$ is the third complex coordinates in the internal space) being invariant 
under the orbifold group. The new graviphoton background interacts with the moduli through both holomorphic and anti-holomorphic parts.
The BRST transformations though modifies only through the holomorphic part of the graviphoton field strength :

\begin{equation}\label{QOmega}
 \begin{aligned}
 & Q_{\Omega}{\chi}=0,~
 Q_{\Omega}{\bar{\chi}}={\eta}~~~~~~&&Q_{\Omega}{\eta}=i\left[{\chi},{\bar{\chi}}\right]\\
 & Q_{\Omega}{a}^{\mu}={M}^{\mu}~&&Q_{\Omega}{M}^{\mu}=i\left[{\chi},{a}^{\mu}\right]-if^{\mu\nu}{a}_{\nu}\\
 & Q_{\Omega}{\lambda}^{c}={D}^{c}~ &&Q_{\Omega}{D}^{c}=i\left[{\chi},{\lambda}^{c}\right]+\epsilon^{cde}{\lambda}_{d}f_{e}\\
 & Q_{\Omega}{w}^{\dot{\alpha}}={\mu}^{\dot{\alpha}}~ &&Q_{\Omega}{\mu}^{\dot{\alpha}}=-i{w}^{\dot{\alpha}}{\chi}+i\phi{w}^{\dot{\alpha}}-\frac{1}{2}f_{\mu\nu}\left(\bar{\sigma}^{\mu\nu}\right)_{\,\,\dot{\beta}}^{\dot{\alpha}}{w}^{\dot{\beta}}\\
 & Q_{\Omega}{\bar{w}}_{\dot{\alpha}}={\bar{\mu}}_{\dot{\alpha}}~~ &&Q_{\Omega}{\bar{\mu}}_{\dot{\alpha}}=i{\chi}{\bar{w}}_{\dot{\alpha}}-i{\bar{w}}^{\dot{\alpha}}\phi-\frac{1}{2}f_{\mu\nu}\left(\bar{\sigma}^{\mu\nu}\right)_{\,\,\dot{\beta}}^{\dot{\alpha}}{\bar{w}}^{\dot{\beta}}\\
 & Q_{\Omega}{\chi}^{\dot\alpha a}={\lambda}^{\dot{\alpha} a}~&& Q_{\Omega}{\lambda}^{\dot{\alpha} a}=i\left[{\chi},{\chi}^{\dot{\alpha} a}\right]-\frac{1}{2}f_{\mu\nu}\left(\bar{\sigma}^{\mu\nu}\right)_{\,\,\dot{\beta}}^{\dot{\alpha}}{\chi}^{\dot{\beta}a}\\
 & Q_{\Omega}{\chi}_{a\dot\alpha}={\lambda}_{a\dot\alpha}~~&& Q_{\Omega}{\lambda}_{a \dot{\alpha}}=i\left[{\chi},{\chi}_{a \dot{\alpha}}\right]-\frac{1}{2}f_{\mu\nu}\left(\bar{\sigma}^{\mu\nu}\right)^{\,\,\dot{\beta}}_{\dot{\alpha}}{\chi}_{a\dot{\beta}}\\
 & Q_{\Omega}{M}^{\alpha a}={C}^{\alpha a}~&&Q_{\Omega}{C}^{\alpha a}=i\left[{\chi},{M}^{\alpha a}\right]-\frac{1}{2}f^{\mu\nu}\left(\sigma_{\mu\nu}\right)^{\alpha}_{\,\,\beta}{M}^{\beta a}\\
 & Q_{\Omega}{\mu}^{a}={h}^{a}~&& Q_{\Omega}{h}^{a}=i{\mu}^{a}{\chi}-i{\phi}{\mu}^{a}\\
 & Q_{\Omega}{\bar{\mu}}^{a}={\bar{h}}^{a}~&&Q_{\Omega}{\bar{h}}^{a}=-i{\chi}{\bar{\mu}}^{a}+i{\bar{\mu}}^{a}{\phi}\\
 \end{aligned}
\end{equation}
which is responsible for the holomorphic part of the gravity-moduli interactions:
\begin{equation}
 Q_{\Omega}{\Xi}=Q{\Xi}+S_{f}
\end{equation}
with $S_{f}$ being interaction terms involving only holomorphic graviphoton field:
\begin{equation}
\begin{split}
 S_{f}=-if^{\mu\nu}{a}_{\nu}\left[{\bar{\chi}},{a}_{\mu}\right]
 -2i{\bar{w}}^{\dot{\alpha}}\phi{w}_{\dot{\alpha}}{\bar{\chi}}
 -if_{\mu\nu}\left(\bar{\sigma}^{\mu\nu}\right)_{\,\,\dot{\beta}}^{\dot{\alpha}}{\bar{w}}^{\dot{\beta}}{w}_{\dot{\alpha}}{\bar{\chi}}
  +2{\bar{\mu}}^{a}{\phi}{\mu}_{a}\\
 -f_{c}{\lambda}^{c}{\eta}
 +\frac{i}{2}\epsilon_{cde}{\lambda}^{c}{\lambda}^{d}f^{e}
 -2if^{\mu\nu}\left(\sigma_{\mu\nu}\right)^{\alpha\beta}{M}^{\alpha a}{M}_{\beta a}\\
 +f_{\mu\nu}\left(\bar{\sigma}^{\mu\nu}\right)_{\dot{\alpha}}^{\,\,\dot{\beta}}{\chi}^{\dot{\alpha}a}\left[{\chi}_{a\dot{\beta}},{\bar{\chi}}\right]
 +if_{c}{D}^{c}{\bar{\chi}}.
 \end{split}
 \end{equation}
The interaction terms involing the anti-holomorphic graviphoton field strength are:
\begin{equation}
\begin{split}
 S_{\bar f}=-f_{c}{\lambda}^{c}{\eta}+if_{c}{D}^{c}{\bar{\chi}}
 +\bar{f}_{\mu\nu}{M}_{\mu}{M}_{\nu}
 +i\bar{f}_{\mu\nu}{a}^{\mu}\left[{\chi},{a}^{\nu}\right]-i\bar{f}_{\mu\nu}{a}^{\mu}f_{\,\,\rho}^{\nu}{a}^{\rho}
 +\bar{f}_{\mu\nu}\left(\bar{\sigma}^{\mu\nu}\right)_{\dot{\alpha}\dot{\beta}}{\bar{\mu}}^{\dot{\alpha}}{\mu}^{\dot{\beta}}\\
 -i\bar{f}_{\mu\nu}\left(\bar{\sigma}^{\mu\nu}\right)_{\dot{\alpha}\dot{\beta}}{\bar{w}}^{\dot{\alpha}}{w}^{\dot{\beta}}{\chi}
 +i\bar{f}_{\mu\nu}\left(\bar{\sigma}^{\mu\nu}\right)_{\dot{\alpha}\dot{\beta}}{\bar{w}}^{\dot{\alpha}}\phi{w}^{\dot{\beta}}
 -\frac{i}{2}\bar{f}_{\mu\nu}\left(\bar{\sigma}^{\mu\nu}\right)_{\dot{\alpha}\dot{\beta}}f_{\rho\sigma}\left(\bar{\sigma}^{\rho\sigma}\right)_{\,\,\dot{\gamma}}^{\dot{\beta}}{\bar{w}}^{\dot{\alpha}}{w}^{\dot{\gamma}}\\
 +\bar{f}^{\mu\nu}\left(\bar{\sigma}_{\mu\nu}\right)^{\dot{\alpha}\dot{\beta}}{\lambda}_{\dot{\alpha}a}{\lambda}_{\dot{\beta}}^{\,\, a}
 +\bar{f}^{\mu\rho}f_{\rho}^{\,\,\nu}\left(\bar{\sigma}_{\mu\nu}\right)_{\dot{\,\,\beta}}^{\dot{\alpha}}{\chi}_{\dot{\alpha}a}{\chi}^{a\dot{\beta}}
 -i\bar{f}^{\mu\nu}\left(\bar{\sigma}_{\mu\nu}\right)_{\,\,\dot{\beta}}^{\dot{\alpha}}{\chi}_{\dot{\alpha}a}\left[{\chi},{\chi}^{a\dot{\beta}}\right]
\end{split}
 \end{equation}
The $S_{\bar f}$ by its own is a $Q_\Omega$-exact expression:
\begin{equation}
 S_{\bar f}=Q_\Omega \Xi_{\bar f}
\end{equation}
where 
\begin{equation}
 \Xi_{\bar f}=if_{c}{\lambda}^{c}{\bar{\chi}}
 +\bar{f}_{\mu\nu}{a}^{\mu}{M}^{\nu}
 +\bar{f}_{\mu\nu}\left(\bar{\sigma}^{\mu\nu}\right)_{\dot{\alpha}\dot{\beta}}{\bar{w}}^{\dot{\alpha}}{\mu}^{\dot{\beta}}
 +\bar{f}^{\mu\nu}\left(\bar{\sigma}_{\mu\nu}\right)_{\,\,\dot{\beta}}^{\dot{\alpha}}{\chi}_{\dot{\alpha}a}{\lambda}^{a\dot{\beta}}.
\end{equation}

So far we have shown that after adding the gravity background to the theory the full action is still a BRST-exact expression where the 
BRST charge and gauge fermion are given by their modified counterparts. We can then write:
\begin{equation}
 S_\Omega=Q_\Omega \Xi_\Omega
\end{equation}
where $S_\Omega$ stands for the full action including the graviphoton interactions, $Q_\Omega$ transformations given in (\ref{QOmega})
and $\Xi_\Omega=\Xi+\Xi_{\bar f}$.
The full action is now expressible in the following form
\begin{equation}
 \begin{split}
  g_{0}^{2}S_{\Omega}=\frac{1}{2}{D}_{c}^{2}
  +\frac{1}{2}{D}_{c}\left(\bar{\eta}_{\mu\nu}^{c}\left[{a}^{\mu},{a}^{\nu}\right]
  +\bar{\zeta}_{mn}^{c}\left[{\chi}^{m},{\chi}^{n}\right]\right)\\
 -\frac{1}{4}\left[{\chi},{\bar{\chi}}\right]\left[{\chi},{\bar{\chi}}\right]
 -\frac{1}{4}\left[{\chi},{\chi}^{m}\right]\left[{\bar{\chi}},{\chi}^{m}\right]
 -\frac{1}{2}\left[{a}_{\mu},{\chi}\right]\left[{a}_{\mu},{\bar{\chi}}\right]\\
 -\frac{1}{2}C^{\alpha a}C_{\alpha a}
 -\frac{1}{2}\left(\bar{\sigma}^{\mu}\right)_{\dot{\alpha}}^{\,\,\alpha}C_{\alpha a}\left[{a}_{\mu},{\chi}^{\dot{\alpha}a}\right]
 -i{\lambda}_{c}Q_\Omega^{2}\left({\lambda}^{c}\right)\\
 -\frac{i}{2}{\eta}\left[{\chi},{\eta}\right]
 -\frac{i}{2}{\lambda}_{\dot{\alpha}a}\left[{\bar{\chi}},{\lambda}^{\dot{\alpha}a}\right]
 -\frac{i}{2}\left(\tau^{c}\right)_{\dot{\alpha}}^{\,\,\dot{\beta}}{\lambda}_{c}\left[{\chi}_{\dot{\beta}a},{\lambda}^{\dot{\alpha}a}\right]\\
 -\frac{i}{2}{\eta}\left[{\chi}_{\dot{\alpha}a},{\lambda}^{\dot{\alpha}a}\right]
 +i{M}_{\mu}\left[{\bar{\chi}},{M}^{\mu}\right]
 -\frac{i}{2}{M}^{\alpha a}Q_\Omega^{2}\left({M}_{\alpha a}\right)\\
 -\frac{i}{2}\left(\sigma^{\mu}\right)_{\alpha}^{\,\,\dot{\alpha}}{M}_{\mu}\left[{\chi}_{\dot{\alpha}a},{M}^{\alpha a}\right]
 +\bar{\eta}_{\mu\nu}^{c}\left[{a}^{\mu},{M}^{\nu}\right]{\lambda}_{c}
 -\frac{i}{2}\left(\bar{\sigma}^{\mu}\right)_{\,\,\alpha}^{\dot{\alpha}}{M}^{\alpha a}\left[{\chi}_{a\dot{\alpha}},{M}_{\mu}\right]\\
 +i\left(\sigma^{\mu}\right)_{\alpha}^{\,\,\dot{\alpha}}\left[{M}^{\alpha a},{a}_{\mu}\right]{\lambda}_{\dot{\alpha}a}
 +2i\left[{a}_{\mu},{M}^{\mu}\right]{\eta}
 -if^{\mu\nu}{a}_{\nu}\left[{\bar{\chi}},{a}_{\mu}\right]\\
 -\left(\tau^{c}\right)_{\,\,\dot{\beta}}^{\dot{\alpha}}f^{c}Q_\Omega^{2}\left({\chi}_{\dot{\alpha}a}\right)\left[{\chi}^{a\dot{\beta}},{\bar{\chi}}\right]
 +2i\left({\bar{\mu}}^{\dot{\beta}}{w}_{\dot{\alpha}}
 +{\bar{w}}_{\dot{\alpha}}{\mu}^{\dot{\beta}}\right)\left(\tau^{c}\right)_{\,\,\dot{\beta}}^{\dot{\alpha}}{\lambda}_{c}\\
 +2i\left({\bar{\mu}}^{\dot{\alpha}}{w}_{\dot{\alpha}}
 +{\bar{w}}_{\dot{\alpha}}{\mu}^{\dot{\alpha}}\right){\eta}
 +2i\left({\bar{\mu}}^{a}{w}_{\dot{\alpha}}
 +{\bar{w}}_{\dot{\alpha}}{\mu}^{a}\right){\lambda}_{\,\, a}^{\dot{\alpha}}\\
 -i{D}^{c}{\bar{w}}^{\dot{\alpha}}{w}_{\dot{\beta}}\left(\tau^{c}\right)_{\,\,\dot{\alpha}}^{\dot{\beta}}
 +2\left({\bar{w}}^{\dot{\alpha}}{h}^{a}
 +{\bar{h}}^{a}{w}^{\dot{\alpha}}\right){\chi}_{a\dot{\alpha}}\\
 -2i{\bar{h}}^{a}{h}_{a}
 +i{\bar{\mu}}^{\dot{\alpha}}{\mu}_{\dot{\alpha}}{\bar{\chi}}
 +i\left({\bar{\mu}}^{\dot{\alpha}}{\mu}^{a}-{\bar{\mu}}^{a}{\mu}^{\dot{\alpha}}\right){\chi}_{\dot{\alpha}a}\\
 -{\bar{w}}^{\dot{\alpha}}Q_\Omega^{2}\left({w}_{\dot{\alpha}}\right){\bar{\chi}}
 +Q_\Omega^{2}\left({\bar{w}}^{\dot{\alpha}}\right){w}_{\dot{\alpha}}{\bar{\chi}}
 +2i{\bar{\mu}}^{i}Q_\Omega^{2}\left({\mu}_{i}\right)\\
 +\bar{f}_{\mu\nu}{M}^{\mu}{M}^{\nu}+\bar{f}^{\mu\nu}{a}_{\mu}Q_\Omega^{2}\left({a}_{\nu}\right)
 +\bar{f}_{\mu\nu}\left(\bar{\sigma}^{\mu\nu}\right)_{\dot{\alpha}\dot{\beta}}{\bar{\mu}}^{\dot{\alpha}}{\mu}^{\dot{\beta}}
 +\bar{f}_{\mu\nu}\left(\bar{\sigma}^{\mu\nu}\right)_{\dot{\alpha}\dot{\beta}}{\bar{w}}^{\dot{\alpha}}Q_\Omega^{2}\left({w}^{\dot{\beta}}\right)\\
 +\bar{f}_{\mu\nu}\left(\bar{\sigma}^{\mu\nu}\right)_{\,\,\dot{\beta}}^{\dot{\alpha}}{\lambda}_{\dot{\alpha}a}{\lambda}^{\dot{\beta}a}+\bar{f}_{\mu\nu}\left(\bar{\sigma}^{\mu\nu}\right)_{\,\,\dot{\beta}}^{\dot{\alpha}}{\chi}_{\dot{\alpha}a}Q_\Omega\left({\chi}^{a\dot{\beta}}\right)
  \end{split}
\end{equation}

\subsection{Localization Limit}\label{local}
The partition function integral will remain unchanged if we rescale the moduli and the anti-holomorphic part of the gravophoton field as:
\begin{equation}
 \begin{split}
& \left({a}^{\mu},{M}^{\mu}\right)\to\frac{1}{x}\left({a}^{\mu},{M}^{\mu}\right)\\
& \left({\bar{\chi}},{\eta}\right)\to\frac{1}{x}\left({\bar{\chi}},{\eta}\right)\\
& \left({w}_{\dot{\alpha}},{\mu}_{\dot{\alpha}}\right)\to\frac{1}{x}\left({w}_{\dot{\alpha}},{\mu}_{\dot{\alpha}}\right)\\
& \left({\chi}_{\dot{\alpha}a},{\lambda}_{\dot{\alpha}a}\right)\to\frac{1}{x}\left({\chi}_{\dot{\alpha}a},{\lambda}_{\dot{\alpha}a}\right)\\
& \left({\lambda}_{c},{D}_{c}\right)\to x^{2}\left({\lambda}_{c},{D}_{c}\right)\\
& \left({M}^{\alpha a},{s}^{\alpha a}\right)\to x^{2}\left({M}^{\alpha a},{s}^{\alpha a}\right)\\
& \left({\mu}^{a},{h}^{a}\right)\to x^{2}\left({\mu}^{a},{h}^{a}\right)\\
& \bar{f}_{\mu\nu}\to z\bar{f}_{\mu\nu}\\
  \end{split}
\end{equation}
In the limit $x\to\infty$ and $\frac{z}{x^{2}}\to\infty$ only few terms in the action survive:
\begin{equation}
 \begin{split}
g_{0}^{2}S_{\Omega}=\frac{1}{2}x^{4}{D}_{c}^{2}-\frac{1}{2}x^{4}C^{\alpha a}C_{\alpha a}-ix^{4}{\lambda}_{c}Q_\Omega^{2}\left({\lambda}^{c}\right)-\frac{i}{2}x^{4}{M}^{\alpha a}Q_\Omega^{2}\left({M}_{\alpha a}\right)\\
 -2ix^{4}{\bar{h}}^{a}{h}_{a}+2ix^{4}{\bar{\mu}}^{a}Q_\Omega^{2}\left({\mu}_{a}\right)+\frac{z}{x^{2}}\bar{f}_{\mu\nu}{M}^{\mu}{M}^{\nu}\\
 +\frac{z}{x^{2}}\bar{f}_{\mu\nu}{a}^{\mu}Q_\Omega^{2}\left({a}^{\nu}\right)+\frac{z}{x^{2}}\bar{f}_{\mu\nu}\left(\bar{\sigma}^{\mu\nu}\right)_{\dot{\alpha}\dot{\beta}}{\bar{\mu}}^{\dot{\alpha}}{\mu}^{\dot{\beta}}\\
 +\frac{z}{x^{2}}\bar{f}_{\mu\nu}\left(\bar{\sigma}^{\mu\nu}\right)_{\dot{\alpha}\dot{\beta}}{\bar{w}}^{\dot{\alpha}}Q_\Omega^{2}\left({w}^{\dot{\beta}}\right)+\frac{z}{x^{2}}\bar{f}_{\mu\nu}\left(\bar{\sigma}^{\mu\nu}\right)_{\,\,\dot{\beta}}^{\dot{\alpha}}{\lambda}_{\dot{\alpha}a}{\lambda}^{\dot{\beta}a}\\
 +\frac{z}{x^{2}}\bar{f}_{\mu\nu}\left(\bar{\sigma}^{\mu\nu}\right)_{\,\,\dot{\beta}}^{\dot{\alpha}}{\chi}_{\dot{\alpha}a}Q_\Omega^{2}\left({\chi}^{\dot{\beta}a}\right)-xf_{c}{\lambda}^{c}{\eta}+ixf_{c}{D}^{c}{\bar{\chi}}+...\\
   \end{split}
\end{equation}
where the dots stand for the terms which are negligible. Note that choosing $f_{c}=f\delta_{c3}$ one can exploit the quartet mechanism
(see e.g. \cite{Billo:2009di}) to integrate out ${\lambda}^{3},{\eta},{D}^{3}$  and ${\bar{\chi}}$ giving an overall factor. 
Then integrating over BRST pairs with index c should be excluded along the null weight of the group representation the pair belongs to.

\section{Weight Vectors of  \texorpdfstring{U$(k)$}{Lg}}\label{WR}
The weight vectors of the group U$(N)$ in terms of the versors $e_i$, $i=1,..,N$ in different representations are given as:\\
 Fundamental representation: 
 \[e_{i}, i=1,...,k\]
 Symmetric two-tensor representation:
 \[2e_{i},\,\,\, e_{i}+e_{j},\,\,\,\, i<j=1,...,k\]
 Antisymmetric two-tensor representation: 
 \[\pm\left(e_{i}+e_{j}\right),\,\, i<j=1,...,k\]
 Adjoint representation: 
 \[\pm\left(e_{i}-e_{j}\right),\,\, k\,\text{times}\,\,0,\,\,\,\,\,\,\, i<j=1,...,k\]
 

\end{document}